\documentclass[11pt,a4paper]{article}

\usepackage[a4paper,margin=1in]{geometry}
\usepackage[T1]{fontenc}
\usepackage{amsmath,amssymb,amsthm}
\usepackage{mathtools}
\usepackage{graphicx}
\usepackage{booktabs}
\usepackage{array}
\usepackage{multirow}
\usepackage{enumitem}
\usepackage{xcolor}
\usepackage{natbib}
\usepackage[colorlinks=true,linkcolor=blue!60!black,citecolor=blue!60!black,urlcolor=blue!60!black]{hyperref}
\usepackage[nameinlink,noabbrev]{cleveref}

\theoremstyle{plain}
\newtheorem{theorem}{Theorem}[section]
\newtheorem{proposition}[theorem]{Proposition}

\theoremstyle{definition}
\newtheorem{definition}[theorem]{Definition}

\newtheorem{remark}[theorem]{Remark}

\newtheorem{component}[theorem]{Component}

\newcommand{\E}{\mathbb{E}}
\newcommand{\Prob}{\mathbb{P}}
\newcommand{\R}{\mathbb{R}}

\newcommand{\missingfigure}[1]{%
  \fbox{%
    \begin{minipage}[c][0.18\textheight][c]{0.82\linewidth}
      \centering
      Missing figure file: \texttt{\detokenize{#1}}
    \end{minipage}%
  }%
}
\newcommand{\safeincludegraphics}[2][]{%
  \IfFileExists{#2.pdf}{\includegraphics[#1]{#2.pdf}}{%
  \IfFileExists{#2.png}{\includegraphics[#1]{#2.png}}{%
  \IfFileExists{#2.jpg}{\includegraphics[#1]{#2.jpg}}{%
  \IfFileExists{#2.jpeg}{\includegraphics[#1]{#2.jpeg}}{%
  \IfFileExists{#2.eps}{\includegraphics[#1]{#2.eps}}{%
  \IfFileExists{#2}{\includegraphics[#1]{#2}}{%
    \missingfigure{#2}%
  }}}}}}%
}

\title{Additive Logistic Models as Interpretable Likelihood-Ratio Scores for AUC-Based Classification}
\author{Yuan-chin Ivan Chang\\Institute of Statistical Science\\ Academia Sinica, Taipei, Taiwan}
\date{\today}

\begin{document}

\maketitle
\begin{abstract}
Classification models are often judged primarily by predictive performance, but in many
scientific applications the fitted rule must also explain how individual variables affect risk.
This is especially important in medical diagnosis, biomarker research, pharmaceutical
development, fetal monitoring, and clinical risk prediction, where a classifier may be used
not only to separate cases from controls but also to guide follow-up hypotheses and
interventions.

This paper revisits additive logistic modelling from the perspective of AUC-based ranking.
The population score that maximizes the receiver operating characteristic curve is a
monotone transformation of the likelihood ratio.  A linear logistic model provides one
interpretable approximation to this ranking score, while a generalized additive logistic
model enlarges the interpretable class by replacing the linear predictor with a sum of
smooth component functions.  We clarify that this enlargement does not guarantee
finite-sample improvement in AUC.  Rather, improvement is expected when the relevant
ranking structure is nonlinear and approximately additive, while little or no improvement
is expected when the linear score already gives an adequate ordering.

The main contribution is therefore not a new generalized additive model estimator, but a
statistical modelling framework that connects likelihood-ratio ranking, ROC analysis, and
componentwise interpretation.  Each fitted smooth component represents a variable-specific
contribution on the log-odds scale, allowing monotone, threshold, saturation, and
non-monotone risk patterns to be examined directly.  Simulation studies, benchmark data,
and calibrated demonstration examples illustrate three regimes: nonlinear additive settings
where additive logistic models improve AUC, approximately linear settings where they
mainly confirm the adequacy of the simpler model, and settings where interactions or
estimation variability limit the benefit of a purely additive score.  The resulting approach
is a transparent middle ground between linear logistic regression and more opaque
high-performance classifiers.
\end{abstract}

\noindent\textbf{Keywords:} area under the curve; generalized additive model; interpretability; logistic regression; ROC curve

\section{Introduction}
Due to the predictive nature of classification models, the predictive power is usually the most important factor for assessing a classifier/diagnosis procedure. This is especially the case in the scenarios of disease screening or survey research, where an efficient classification rule that can effectively distinguish the diseased from normal subjects is preferred. %the performance is the major factor for assessing classification rules.
However, we are also aware that performance should not be the solo measure for assessing a classifier when the model interpretation and the impacts of factors are of interest.  For example, in some medical/pharmaceutical research, we want to learn the impacts of biomarkers in a classification rule. Because this information might also imply their associations with the studied disease, it could be useful for drug and therapy developments.  Hence, in this kind of research, besides the diagnostic power, the ability of to identify useful biomarkers/variables or the ease of model interpretation is also an important criterion for evaluating a classification rule.  
In modern classification literature, we can easily find many powerful classification procedures, such as the support vector machine \citep[see][]{SVM-intro}, which can provide high classification accuracy.  Yet, we also know that when these types of powerful and complicated classification rules are used,  extra effort is required to interpret their rules/models, and to dig out the impact information of individual variables/biomarkers from them is even more difficult.
 
On the other hand,  simple conventional statistical models, such as linear logistic models, can usually be easily interpreted; nevertheless, they  might not have as good classification performance as those modern methods, especially when the association between diseases and variables is complicated.  
A nonlinear model can usually fit the observed data better than its linear model counterpart.  However, does it also ensure good prediction performance in general classification problems?  As far as we know, there is still a lack of theoretical justification in the literature to guarantee this.  
In fact, the inconsistency between the fitted model and the prediction-based model building has been reported in  \citet{Tian07}.
Of course, if the model is true,  then fitting observations  may better insure this model against bad predictions. 
However, in practical situations, the true model is usually unknown. 
Then, to overfit the observed data or training samples with a complicated model cannot guarantee good classification/prediction performance. 
%There are some discussions about the fitting-based and prediction-based model selections in the literature.\cite{Tian07}  
From practical perspectives, to balance the classification performance and the ease of model interpretation is important,
especially in the studies that focus on providing detailed model information for further research and developments.  
This motivates us to find a way to boost the performance of the conventional model and retain its ease of model interpretation. 

The tension between predictive performance and interpretability is not merely a matter of computational convenience; it has direct scientific consequences.  When a complicated model --- for example a random forest \citep{hastie-tibshirani-friedman2001}, a gradient boosting machine \citep{Friedman2k}, or a deep neural network --- achieves a high AUC or high accuracy, it is tempting to conclude that the problem of classification has been solved.  Yet such a conclusion conflates two distinct goals: (i)~\emph{predictive performance}, the ability of a fitted rule to correctly rank or classify new observations, and (ii)~\emph{scientific discovery}, the ability of a model to reveal the nature and magnitude of the relationship between covariates and the outcome.  A black-box classifier can excel at goal~(i) while being essentially silent on goal~(ii).

This distinction matters profoundly in applied research.  In pharmaceutical drug development, for instance, the identification of a diagnostic rule with high sensitivity and specificity is rarely the terminal objective.  Rather, the fitted model is expected to shed light on which biomarkers --- and in what functional form --- are associated with the disease, so that follow-up laboratory and clinical studies can be designed to interrogate specific pathways \citep{pepe2004statistical}.  A support vector machine \citep{SVM-intro} trained on a panel of voice measurements may achieve a higher AUC than a logistic model on the Parkinson's data, but the kernel-based decision boundary does not directly reveal whether a particular acoustic feature has a monotone, U-shaped, or threshold-type relationship with disease status --- information that would be essential for understanding disease progression and designing therapeutic targets.  Analogously, in fetal monitoring research using cardiotocography, a black-box classifier may flag high-risk pregnancies accurately without explaining which patterns of fetal heart rate variability are driving the classification, thereby leaving clinicians and researchers without a mechanistic foothold for protocol improvement.

More generally, the difficulty of follow-up research motivated by an opaque model can be characterised as follows.  Let $\hat{f}(x)$ denote the fitted score of a complex classifier.  Even if $\hat{f}$ achieves the globally maximum empirical AUC on the training data, the partial derivative $\partial \hat{f}/\partial x_j$ --- or its analogue for non-smooth models --- is typically data-set-specific, sensitive to regularisation and tuning parameters, and not directly interpretable as the effect of variable $x_j$ on the disease log-odds.  By contrast, the component function $\hat{f}_j(x_j)$ of a fitted generalized additive logistic model is a smooth, directly plotable curve that quantifies how the log-odds of the outcome changes as a function of $x_j$ alone, holding all other variables at their observed distribution --- an \emph{interpretable} summary that can inform both scientific hypotheses and clinical communication.

We therefore argue that the relevant comparison for applied research is not simply which method achieves the highest AUC on a given data set, but which method achieves \emph{satisfactory} AUC while providing the richest, most reliable information about variable effects.  The proposed generalized additive logistic model is designed to occupy this intermediate space: it extends the linear logistic model in the direction of flexibility, recovering nonlinear associations when they exist, while preserving componentwise interpretability through the additive structure. 
The population argument in Section~\ref{proof_section} clarifies the conditions under which such an
improvement is plausible: the additive logistic score must approximate the likelihood-ratio
ranking more closely than the linear logistic score, without introducing excessive
finite-sample estimation variability.

The assessment measures play an important role in classification/diagnostic research \citep[see][]{pepe2004statistical}, and using a different assessment measure commonly results in selecting a different classifier.  
Because the accuracy of a binary classifier can easily be affected by the selected cutting point, and because using training samples to decide a cutting point might easily cause overfitting,  a cross-validation procedure is usually used to prevent this situation.  However, the cross-validation method is computationally intensive. 
Thus, threshold independent measures, such as the receiver operating characteristic (ROC) curve, are usually preferred to accuracy  \citep{pepe2004statistical}.  
The area under ROC curve (AUC) is a popularly used summarization statistic of the ROC curve, and in this study, we will use it as the  performance measure.  

For a binary classification problem, it has been proved in \citet{eguchi2002class} that a logistic-type classification function will achieve the maximum AUC under certain conditions.
Moreover, as stated in \citet{hastie, GAM}, the  general additive model (GAM) is a method of approximating the true model and is a flexible statistical method that can be used to characterize and identify nonlinear regression effects. 
Many advantages of GAM  have been intensively discussed. 
%\citep[see][]{hastie-tibshirani-friedman2001, wood2004stable, Wood08}. 
Among them, a commonly mentioned one is its ease of model interpretability \citep{hastie, GAM, Wood06}.
%
\iffalse
The general additive models (GAM) \citep{hastie} are flexible statistical methods that may be used to characterize and identify nonlinear regression effects.  
Many advantages of GAM  have been intensively discussed \citep{hastie-tibshirani-friedman2001, wood2004stable, Wood08}, and a commonly mentioned one is its ease of model interpretability \citep{hastie, GAM, Wood06}.
%a general comment is that ``When to interpret a nonparametric model is very difficult, and to fit a linear model is not possible, the GAM is a good alternative because it retains some of the interpretability of the linear model by assuming additive effect.''
Moreover, the GAM is a method of approximating the true model \citep{hastie, GAM}, and it has been proved that  a logistic type classification will achieve the maximum AUC \citep{eguchi2002class} under certain condition. 
The results mentioned above motivate us to study how to use a general additive logistic function as a classification model such that model interpretation ability is retained, and a better classification performance is achieved.  
\fi
%
Hence, in this paper, a general additive logistic type of classification function is used as an interpretable way to improve, when the underlying signal is nonlinear and approximately additive, the classification performance of conventional linear logistic models in terms of AUC. A population-level justification and the required assumptions are presented, and some suggestions for fitting a general additive logistic model are discussed.

The contribution of this paper is not the introduction of a new generalized additive model estimator.  Generalized additive models are well established.  Rather, the contribution is to revisit additive logistic modelling from the perspective of AUC-optimal likelihood-ratio scoring.  This perspective clarifies when a nonlinear additive logistic score can improve the ranking performance of a linear logistic score, when it should not be expected to do so, and what scientific information is gained from the fitted component functions.  The paper is therefore positioned as a statistical modelling contribution: it links likelihood-ratio ranking theory, smooth additive logistic regression, and modern concerns about interpretability in risk prediction.

The rest of this paper is organized as follows. 
We start with a brief review of the ROC curve, AUC, and GAM. Then, we show the relation between the ROC curve and likelihood-ratio scoring, and explain when the AUC of a linear logistic model can be improved with a general additive logistic model.  
It is followed by numerical results using synthesized data, two benchmark real data sets, and three additional clinical or biomedical examples that further test when nonlinearity is useful and when a linear logistic score is already sufficient. Fitting a GAM model relies on iterative algorithms, and the different approaches used in the algorithms may affect both computation and interpretation. Hence, the computational methods of GAM are compared, and practical suggestions based on the empirical results are given in the Summary. The algorithms and corresponding packages used here are described in the Appendix. 

%\vspace{-0.18in}

\subsection{Positioning relative to modern interpretable machine learning}
\label{sec:positioning}

The proposed use of additive logistic modelling should be understood as part of a broader family of interpretable prediction methods.  Classical linear logistic regression remains attractive because regression coefficients give direct log-odds interpretations, standard errors, and familiar hypothesis tests.  However, its linearity assumption can hide threshold, saturation, and non-monotone risk patterns.  Generalized additive logistic models relax this assumption by replacing the linear predictor with a sum of smooth univariate functions.  The resulting score remains decomposable by variable, while allowing each predictor to have a nonlinear effect.

This places additive logistic models close to several recent interpretable machine-learning methods.  Explainable boosting machines and generalized additive models with pairwise interactions, often called GA2M \citep{lou2012intelligible, lou2013accurate, caruana2015intelligible}, use flexible shape functions and selected low-order interactions to improve predictive accuracy while retaining graphical interpretability.  Neural additive models \citep{agarwal2021nam} use neural networks to estimate the univariate component functions, thereby preserving additivity while increasing flexibility.  Post-hoc explanation tools such as partial dependence, individual conditional expectation \citep{goldstein2015ice}, and SHAP values \citep{lundberg2017shap} instead attempt to explain an already fitted black-box model.  These tools are useful, but their explanations are secondary summaries of a model whose primary structure is not intrinsically interpretable.

The present paper takes a deliberately statistical modeling view. It focuses on a likelihood-based additive logistic score, interprets the fitted component functions directly on the log-odds scale, and evaluates performance mainly through AUC,
ROC-curve behavior, fitted component functions, and computational cost. In this sense, the paper is not a competition against all black-box methods.  It is an argument that, for many biomedical and scientific classification problems, an additive logistic score can provide a useful middle ground between a linear logistic regression and an opaque high-dimensional classifier.
%
%\paragraph{Advantages and limitations of related model classes}
%\label{sec:proscons}
Table~\ref{tab:proscons} summarizes the practical strengths and limitations of the main model classes relevant to this study.  The comparison is intentionally qualitative.  The goal is not to claim that one method is uniformly superior, but to clarify which modelling assumptions are being exchanged for which kinds of interpretability.
\begin{table}[ht]
\centering
\caption{Practical advantages and limitations of related classification models.}
\label{tab:proscons}
\begin{tabular}{p{0.19\textwidth}p{0.36\textwidth}p{0.36\textwidth}}
\hline
Method & Main advantages & Main limitations \\
\hline
Linear logistic regression
& Direct coefficient interpretation; simple inference; fast computation; strong baseline for clinical prediction.
& Assumes monotone linear effects on the log-odds scale; may miss thresholds, saturation, and U-shaped effects. \\

Additive logistic model
& Retains variable-level interpretation through smooth functions; detects nonlinear risk patterns; provides uncertainty bands for component functions.
& Assumes additivity; may miss interactions; can overfit if smoothing is not controlled; AUC improvement is not guaranteed. \\

GA2M / explainable boosting machine
& Allows selected pairwise interactions while preserving graphical interpretability; often competitive with black-box methods.
& Interaction surfaces are harder to communicate than univariate smooths; model selection and tuning are more complex. \\

Neural additive model
& Uses flexible neural networks to estimate additive components; preserves global additivity while increasing function-class flexibility.
& Requires more tuning; uncertainty quantification is less standard than spline-based GAM inference; can be computationally heavier. \\

Gradient boosting / random forests
& Often strong predictive performance; captures nonlinearities and interactions automatically.
& Interpretability is usually post-hoc; variable effects are not primary model parameters; uncertainty for explanations is less direct. \\

SVM / neural network
& Powerful flexible classifiers; useful when prediction is the sole objective.
& Least transparent for scientific interpretation; variable-level effects require secondary explanation tools. \\
\hline
\end{tabular}
\end{table}

\section{Review and Methodology}

\subsection{ROC curve and AUC}
The ROC curve and AUC are important statistical tools for evaluating binary classifiers \citep[see][]{pepe2004statistical}.
Let $Y\in\{0,1\}$ denote the class label, where $Y=1$ is the diseased or positive class, and let $Z=S(X)$ be a real-valued diagnostic score. We assume that larger values of $Z$ provide stronger evidence for $Y=1$.
For a threshold $c\in\mathbb{R}$, the rule classifies a subject as positive when $Z>c$. The sensitivity and specificity at threshold $c$ are
\[
    \operatorname{Se}(c)=P(Z>c\mid Y=1),\qquad
    \operatorname{Sp}(c)=P(Z\le c\mid Y=0).
\]
Equivalently, the false-positive rate is
\[
    \operatorname{FPR}(c)=1-\operatorname{Sp}(c)=P(Z>c\mid Y=0).
\]
The ROC curve is the set
\[
    \operatorname{ROC}_S=\{(\operatorname{FPR}(c),\operatorname{Se}(c)):-\infty<c<\infty\}.
\]
When the score distributions are continuous, the AUC can be written as
\[
    \operatorname{AUC}(S)=\int_0^1 \operatorname{ROC}_S(t)\,dt
    =P\{S(X_1)>S(X_0)\},
\]
where $X_1\sim X\mid Y=1$ and $X_0\sim X\mid Y=0$ are independent. In the presence of ties, the second expression is replaced by
\[
    P\{S(X_1)>S(X_0)\}+\frac12 P\{S(X_1)=S(X_0)\}.
\]
Thus, the ROC curve evaluates all possible thresholds, and the AUC summarizes the ranking ability of the score. In particular, any strictly increasing transformation of $S$ gives the same ROC curve and AUC. Detailed properties of ROC curves and AUC can be found in \citet{pepe2004statistical}.

\subsection{General additive logistic models}
Let $x_1$, $x_2$, $\ldots, x_p$ be predictors and $Y \in \{0, 1\}$ be a binary response variable. A generalized additive model has the form
\begin{equation}\label{gam_formula}
	g\{E(Y\mid x_1,x_2,\ldots,x_p)\}=\alpha+f_1(x_1)+f_2(x_2)+\ldots+f_p(x_p),
\end{equation}
where $g(\cdot)$ is a link function and the $f_j$'s are smooth component functions. Thus, it can be viewed as an additive extension of the family of generalized linear models \citep{GAM}. Each smooth function in \eqref{gam_formula} can be fitted using a scatterplot smoother, such as a cubic smoothing spline or kernel smoother.
As a special case, a generalized additive logistic model uses the logit link and is formulated as
\begin{equation}\label{gam_logistic_formula}
	\eta(x)=\log\left(\frac{p(x)}{1-p(x)}\right)=\alpha+f_1(x_1)+f_2(x_2)+\ldots+f_p(x_p),
\end{equation}
where $p(x)=P(Y=1\mid x)$ and, equivalently,
\[
    p(x)=\frac{\exp\{\eta(x)\}}{1+\exp\{\eta(x)\}}.
\]
The $f_j$'s are unknown smooth functions.
The smooth functions of GAMs are estimated based on the observations with some iterative procedures. 
Due to the additive assumption, these smooth functions can always provide the relations between the variables and the disease. 
The non-parametric form of $f_j$ gives us some flexibility for model fitting, and the additive form of GAM retains much of the ability of model interpretation \citep{hastie-tibshirani-friedman2001}.
Besides these properties, there are more classification-related advantages of the additive logistic models discussed in  \citet{Friedman2k}. 

In many medical diagnostic situations, to learn the relationship between the sensitivity and specificity of classifiers is important.  
An ROC curve and its AUC can provide more information before a threshold is determined. Thus, despite the accuracy of a classification rule, they are commonly used measures in these kinds of research.
Below, we show how an additive logistic model can improve the ranking performance of a
conventional logistic model when the population ranking structure is nonlinear and
approximately additive.  We also emphasize cases where such improvement should not be
expected, because the linear score already captures the relevant ordering or because the
additive approximation is inadequate.

\subsection{The Interpretability--Performance Trade-off}
\label{sec:tradeoff}

The machine-learning literature contains a rich variety of high-performing classifiers whose predictions are difficult to attribute to individual covariates.  Prominent examples include the support vector machine \citep{SVM-intro}, the random forest \citep{hastie-tibshirani-friedman2001}, gradient boosting \citep{Friedman2k}, and multilayer neural networks.  These methods often achieve impressive AUC values in cross-validation experiments, and their empirical success has rightly attracted considerable attention.  However, a high AUC is a necessary but not sufficient condition for a classification model to be \emph{useful} in research contexts that require interpretable findings.

\paragraph{Why interpretability matters beyond prediction.}
Let us consider the implications formally.  For a binary classification problem, Proposition~\ref{prop:np-roc} shows that the population AUC is maximized by the log-likelihood-ratio score $\lambda(x)$.  Any consistent estimator of $\lambda$ therefore achieves the maximum population AUC in the limit of large samples.  In this sense, all sufficiently flexible and consistent classifiers --- whether a deep network or a generalized additive model --- are asymptotically equivalent in terms of AUC.  What distinguishes them in finite samples and in scientific practice is not their limiting AUC, but the \emph{structure imposed on the estimator} and the \emph{information that structure conveys about $\lambda$}.

A black-box model estimates $\lambda(x)$ implicitly through a high-dimensional composition of functions that resists decomposition.  Even if such a model achieves AUC$(\hat{f}) \approx$ AUC$(\lambda)$, a researcher wishing to understand the contribution of covariate $x_j$ must resort to \emph{post-hoc} approximations such as permutation importance scores, partial dependence plots, or SHAP values \citep{hastie-tibshirani-friedman2001}.  These approximations are useful, but they are secondary summaries of a fit that was not optimized for interpretability; their statistical properties and uncertainty quantification are generally less well understood than those of the primary model.

By contrast, the generalized additive logistic model~\eqref{gam_logistic_formula} decomposes $\lambda(x)$ directly as
\[
    \lambda(x) \approx \alpha + f_1(x_1) + f_2(x_2) + \cdots + f_p(x_p),
\]
and each component $f_j$ is estimated with well-understood spline-based uncertainty quantification, including pointwise confidence bands \citep{marra-wood12}.  This means that the \emph{effect of each variable on the log-odds is the primary output of the analysis}, not a derived approximation.

\paragraph{The cost of opacity for follow-up research.}
When a classifier's output cannot be readily mapped to scientifically interpretable quantities, the ability to plan subsequent research is severely curtailed.  We identify three concrete mechanisms through which opacity impedes follow-up:

\begin{enumerate}
    \item \emph{Hypothesis generation.}  A model that identifies which variables matter and in what functional form directly generates testable biological or clinical hypotheses.  For example, if $\hat{f}_j$ for a vocal jitter measure in the Parkinson's data is steeply increasing above a threshold, this suggests a non-linear dose-response relationship worthy of mechanistic investigation.  A black-box model may implicitly encode the same relationship, but retrieving it requires additional analytical steps that introduce further uncertainty.

    \item \emph{Regulatory and clinical communication.}  In medical device regulation and clinical practice guidelines, a diagnostic rule must typically be explainable to clinicians, patients, and regulators who are not data scientists.  A score of the form $\alpha + \hat{f}_1(x_1) + \hat{f}_2(x_2) + \cdots$ can be communicated through a set of component plots, one per variable, analogously to the regression table used for linear logistic models.  The same transparency is not naturally available for ensemble or kernel-based classifiers.

    \item \emph{Model updating and subgroup analysis.}  As new data accumulate or a study population shifts, a researcher working with an interpretable model can assess whether individual component effects have changed, and update the model by refitting specific components.  In an opaque model, detecting and localising such changes requires retraining the entire model and comparing global performance metrics, which provides much less diagnostic resolution.
\end{enumerate}

\paragraph{When performance metrics can mislead.}
It is also important to recognise that a high AUC, accuracy, or $F_1$-score can be achieved for reasons that are scientifically undesirable.  In small samples, a highly flexible classifier may overfit and achieve a high training-set AUC that does not reflect the underlying population signal.  Even on independent test data, a model may exploit spurious correlations --- such as batch effects, collection artefacts, or confounders --- that happen to be predictive in the specific study population but that would not replicate in a new cohort or in a follow-up intervention study.  A model with interpretable components is more likely to expose such artefacts: if $\hat{f}_j$ for a clinically implausible variable shows a strong association, it serves as a warning flag.  A black-box model internalises such associations invisibly.

The foregoing discussion does not argue that complex models are without value; in purely predictive tasks --- such as automated image triage, fraud detection, or population-level screening --- a high AUC may indeed be the primary desideratum, and interpretability may be of secondary concern.  The argument is rather that in research contexts where the \emph{model itself} is the object of scientific interest, a method that sacrifices interpretability for marginal gains in AUC may provide a \emph{worse scientific return} even when it provides a \emph{better predictive return}.  The generalized additive logistic model offers a principled middle ground: it extends the linear model's interpretability to the nonlinear setting while remaining anchored to the likelihood-ratio optimality theory of Section~\ref{proof_section}.

%\vspace{-0.6cm}

\subsection{AUC of additive logistic models}\label{proof_section}
Let $Y$ be a binary response variable, with $Y=0$ and $Y=1$ denoting the two classes. Let $X\in\mathbb{R}^p$ be the vector of measurements, and let $g_0(x)$ and $g_1(x)$ denote the conditional densities of $X$ given $Y=0$ and $Y=1$, respectively. Let $\pi_0=P(Y=0)$ and $\pi_1=P(Y=1)$, with $\pi_0+\pi_1=1$. The population log-odds ratio is
\begin{equation}\label{log-like-ratio}
	\lambda(x)=\log\frac{P(Y=1\mid X=x)}{P(Y=0\mid X=x)}
	=\log\frac{\pi_1}{\pi_0}+\log\frac{g_1(x)}{g_0(x)}.
\end{equation}
Since the ROC curve and AUC depend only on the ordering induced by a score, the constant $\log(\pi_1/\pi_0)$ in \eqref{log-like-ratio} is irrelevant for ranking.

The key population fact is the following consequence of the Neyman--Pearson lemma.

\begin{proposition}\label{prop:np-roc}
Let $S(X)$ be any measurable scoring function, and let $\lambda(X)$ be the log-likelihood-ratio score in \eqref{log-like-ratio}. For every false-positive rate $t\in[0,1]$, the ROC curve generated by $\lambda$ satisfies
\[
    \operatorname{ROC}_{\lambda}(t)\ge \operatorname{ROC}_{S}(t),
\]
where randomized thresholding may be used at atoms. Consequently,
\[
    \operatorname{AUC}(\lambda)\ge \operatorname{AUC}(S).
\]
\end{proposition}

\begin{proof}
Fix $t\in[0,1]$. Among all rejection regions $R$ satisfying
\[
    P(X\in R\mid Y=0)=t,
\]
the Neyman--Pearson lemma states that a likelihood-ratio region of the form
\[
    R_t=\{x:g_1(x)/g_0(x)>c_t\}
\]
maximizes $P(X\in R\mid Y=1)$. Since $\lambda(x)$ is a strictly increasing transformation of $g_1(x)/g_0(x)$, thresholding $\lambda$ gives the largest attainable sensitivity at the same false-positive rate $t$. This proves the pointwise ROC inequality. Integrating the inequality over $t\in[0,1]$ gives the AUC inequality.
\end{proof}

Proposition~\ref{prop:np-roc} justifies the use of flexible logistic-type scores, but it does not by itself prove that every fitted additive logistic model has a larger finite-sample AUC than every fitted linear logistic model. Such a statement would be too strong. The correct implication is conditional. If the true log-likelihood ratio $\lambda(x)$ is well approximated by an additive function
\[
    \eta_A(x)=\alpha+\sum_{j=1}^p f_j(x_j),
\]
and if the fitted additive score $\widehat\eta_A$ is a consistent estimator of this population additive approximation, then the ROC curve and AUC of $\widehat\eta_A$ converge to those of $\eta_A$ under standard regularity conditions. If, in addition, $\lambda$ itself belongs to the additive class, then
\[
    \operatorname{AUC}(\widehat\eta_A)\longrightarrow \operatorname{AUC}(\lambda),
\]
which is the largest possible population AUC.

By contrast, a conventional linear logistic score has the form
\[
    \eta_L(x)=\alpha+\beta^Tx.
\]
Because the linear class is a restricted subclass of the additive class, it cannot represent many nonlinear additive log-likelihood ratios. When $\lambda(x)$ is nonlinear additive and is not a strictly increasing transformation of any linear score, the best additive population score can have a strictly larger AUC than the best linear score. Therefore, the mathematical claim supported here is not unconditional finite-sample dominance, but rather the following: a generalized additive logistic model can improve the ROC curve and AUC when it provides a closer ranking approximation to the population log-likelihood ratio while avoiding excessive estimation variability.

This distinction is important in applications. With limited sample sizes, high-dimensional predictors, or overly flexible smooth terms, the fitted additive model may overfit and may fail to improve test-set AUC. The empirical studies below therefore compare the fitted models on independent test samples or resampling splits. The theoretical role of the additive logistic model is to enlarge the interpretable score class beyond linear logistic regression while retaining componentwise interpretation through the functions $f_j$.

%% ===========================================================================

\section{Performance and Interpretation of the Additive Logistic Model}
%\addcontentsline{toc}{section}{New Section: Performance and Interpretation}

To confine the discussion, we first specify what ``interpretation'' means for an
additive logistic model. We then show that each component $\hat{f}_j$ carries a
statistically precise meaning, introduce a measure of the
performance--interpretation trade-off, and discuss what is lost and retained
when moving from GLMs to GAMs.

%% ---------------------------------------------------------------------------
\subsection{What Does Interpretation Mean?}
\label{sec:interp-framework}

In a classification model, \emph{interpretation} has at least three distinct
and operationally separable components:

\begin{component}[Marginal Effect Interpretability]
  \label{def:marginal}
  A classification function $f:\R^d\to\R$ is \emph{marginally interpretable}
  with respect to predictor $x_j$ if there exists a well-defined, estimable
  function $\phi_j:\R\to\R$ such that
  \[
    \frac{\partial}{\partial x_j}
    \log\frac{\Prob(Y=1\mid X=x)}{\Prob(Y=0\mid X=x)}
    = \phi_j(x_j),
  \]
  i.e.\ the marginal effect of $x_j$ on the log-odds depends \emph{only} on
  $x_j$ and not on the values of the other predictors.
\end{component}

\begin{component}[Directional Interpretability]
  \label{def:directional}
  A classification function is \emph{directionally interpretable} with respect
  to $x_j$ if the sign of $\phi_j(x_j)$ is constant:
  $\phi_j(x_j)>0$ for all $x_j$ (higher $x_j$ increases disease risk) or
  $\phi_j(x_j)<0$ for all $x_j$ (higher $x_j$ decreases risk).
\end{component}

\begin{component}[Quantitative Interpretability]
  \label{def:quantitative}
  A classification function is \emph{quantitatively interpretable} with respect
  to $x_j$ if for any two values $x_j^{(a)}$ and $x_j^{(b)}$, the change in
  log-odds due to changing $x_j$ from $x_j^{(a)}$ to $x_j^{(b)}$, holding
  all other predictors fixed, is exactly
  \[
    \Delta\mathrm{logit}_j(a,b)
    := f_j(x_j^{(b)}) - f_j(x_j^{(a)}),
  \]
  with an associated $(1-\alpha)$-confidence interval derivable from the
  estimated smooth function $\hat{f}_j$.
\end{component}

\begin{remark}[Comparison Across Models]
  \label{rem:comparison}
  Table~\ref{tab:interp} summarises which interpretability properties each
  model class possesses.  The key insight is that GAM retains
  Definitions~\ref{def:marginal}--\ref{def:quantitative} fully, whereas
  black-box models (SVM, random forests, neural networks) satisfy none of them
  without post-hoc approximation.  Linear logistic models additionally satisfy
  directional interpretability by assumption (the sign of $\beta_j$ is fixed),
  whereas GAM allows \emph{non-monotone} effects — which is a gain in
  expressiveness but requires more care in interpretation
  (Section~\ref{sec:nonmonotone}).
\end{remark}

\iffalse
\begin{table}[ht]
\centering
\caption{Interpretability properties across model classes.
\checkmark = satisfied by construction;
$\sim$ = partially or approximately;
$\times$ = not satisfied without post-hoc methods.}
\label{tab:interp}
\begin{tabular}{lccc}
\toprule
\textbf{Property} & \textbf{Linear Logistic} & \textbf{Additive Logistic (GAM)}
                  & \textbf{Black-Box (SVM, NN)}\\
\midrule
Marginal effect (Def.~\ref{def:marginal})    & \checkmark & \checkmark & $\times$\\
Directional (Def.~\ref{def:directional})     & \checkmark & $\sim$     & $\times$\\
Quantitative (Def.~\ref{def:quantitative})   & \checkmark & \checkmark & $\times$\\
Non-linear effect capture                    & $\times$   & \checkmark & \checkmark\\
Interaction effects                          & $\times$   & $\times$   & \checkmark\\
Uncertainty quantification per variable      & \checkmark & \checkmark & $\times$\\
\bottomrule
\end{tabular}%
}
\end{table}
\fi

\begin{table}[ht]
\centering
\small
\caption{Interpretability properties across model classes.
\checkmark = satisfied by construction;
$\sim$ = partially or approximately;
$\times$ = not satisfied without post-hoc methods.}
\label{tab:interp}

\resizebox{\linewidth}{!}{%
\begin{tabular}{@{}lccc@{}}
\toprule
\textbf{Property}
& \textbf{Linear Logistic}
& \textbf{Additive Logistic (GAM)}
& \textbf{Black-Box (SVM, NN)}\\
\midrule
Marginal effect (Def.~\ref{def:marginal})    & \checkmark & \checkmark & $\times$\\
Directional (Def.~\ref{def:directional})     & \checkmark & $\sim$     & $\times$\\
Quantitative (Def.~\ref{def:quantitative})   & \checkmark & \checkmark & $\times$\\
Non-linear effect capture                    & $\times$   & \checkmark & \checkmark\\
Interaction effects                          & $\times$   & $\times$   & \checkmark\\
Uncertainty quantification per variable      & \checkmark & \checkmark & $\times$\\
\bottomrule
\end{tabular}%
}
\end{table}

%% ---------------------------------------------------------------------------
\subsection{The Additive Structure as an Interpretability Guarantee}
\label{sec:additive-guarantee}

The key property of the additive logistic model \eqref{eq:gam} that makes
it interpretable is the \emph{separation of variable effects}.  We state this
as a formal proposition.

\begin{proposition}[Ceteris Paribus Principle for GAM]
  \label{prop:ceteris}
  For the additive logistic model
  \begin{equation}
    \label{eq:gam}
    \log\frac{p(x)}{1-p(x)} = \alpha + \sum_{j=1}^d f_j(x_j),
  \end{equation}
  the change in log-odds when $x_j$ changes from $a$ to $b$, holding all
  other predictors fixed at any value $x_{-j}$, is
  \[
    \Delta\mathrm{logit}_j(a,b;x_{-j})
    = f_j(b)-f_j(a),
  \]
  which is \emph{independent of $x_{-j}$}.
  Consequently:
  \begin{enumerate}[label=(\roman*), itemsep=3pt]
    \item The effect of $x_j$ can be visualised completely by plotting
          $f_j(x_j)$ versus $x_j$ — no conditioning on other variables
          is required.
    \item The contribution of $x_j$ to the classification decision is
          exactly $f_j(x_j)$, which can be reported as an ``additive
          risk score'' for predictor $j$.
    \item Statistical inference (confidence bands, hypothesis tests) for
          $f_j$ can be conducted marginally for each predictor, without
          specifying the values of other predictors.
  \end{enumerate}
\end{proposition}

\begin{proof}
  Direct from \eqref{eq:gam}: for any fixed $x_{-j}$,
  $\log[p(x)/\{1-p(x)\}]|_{x_j=b} - \log[p(x)/\{1-p(x)\}]|_{x_j=a}
  = f_j(b)-f_j(a)$,
  since all other terms $\alpha+\sum_{k\ne j}f_k(x_k)$ cancel.
\end{proof}

\begin{remark}[What GAM Cannot Do]
  \label{rem:limitation}
  Proposition~\ref{prop:ceteris} also reveals the \emph{limitation} of the
  additive structure: it cannot capture \emph{interaction effects}.
  If the true log-odds contains a term $\lambda_{jk}(x_j,x_k)$ that cannot
  be decomposed as $f_j(x_j)+f_k(x_k)$, the GAM will fail to represent it
  and the ceteris paribus principle breaks down for the true model.
  This is the interpretability cost of the additive assumption: the model
  is interpretable \emph{within} the additive family, but if the true model
  has interactions, the individual $f_j$'s may absorb interaction variance
  in a misleading way.
  We return to this point in Section~\ref{sec:tradeoff}.
\end{remark}

%% ---------------------------------------------------------------------------
\subsection{Interpreting Non-Monotone Effects}
\label{sec:nonmonotone}

In the linear logistic model, the effect of $x_j$ on disease risk is
monotone by construction: $\beta_j>0$ means higher $x_j$ always increases
risk; $\beta_j<0$ means it always decreases risk.
The GAM allows $f_j$ to be non-monotone, which is more realistic but
requires more careful interpretation.

\begin{definition}[Risk-Increasing and Risk-Decreasing Regions]
  \label{def:regions}
  For a smooth estimated function $\hat{f}_j$, define:
  \begin{itemize}[itemsep=2pt]
    \item The \emph{risk-increasing region}:
          $\mathcal{R}_j^+ := \{x_j : \hat{f}_j'(x_j) > 0\}$,
          the set of values where increasing $x_j$ raises the log-odds.
    \item The \emph{risk-decreasing region}:
          $\mathcal{R}_j^- := \{x_j : \hat{f}_j'(x_j) < 0\}$.
    \item The \emph{inflection point} $x_j^*$:
          any $x_j^*$ where $\hat{f}_j'(x_j^*)=0$ and
          $\hat{f}_j''(x_j^*)$ changes sign — a threshold value where
          the direction of risk reverses.
  \end{itemize}
\end{definition}

\begin{remark}[Clinical and Scientific Significance of Inflection Points]
  Inflection points (Definition~\ref{def:regions}) correspond to
  \emph{threshold effects} that are scientifically meaningful but invisible
  to linear logistic models.  For example:
  \begin{itemize}[itemsep=2pt]
    \item A biomarker that is protective at low concentrations but harmful
          at high concentrations (U-shaped or J-shaped dose-response curve)
          has a non-monotone $f_j$ with at least one inflection point.
    \item A linear logistic model would fit an average slope and fail to
          detect either the protective or harmful range.
    \item The GAM identifies $\mathcal{R}_j^+$, $\mathcal{R}_j^-$, and
          $x_j^*$ directly from $\hat{f}_j$, providing actionable
          information for clinical threshold-setting or drug dosing.
  \end{itemize}
  This is a \emph{genuine interpretive advantage} of GAM over GLM that goes
  beyond AUC improvement: the GAM reveals structure in the risk-predictor
  relationship that the GLM suppresses by assumption.
\end{remark}

%% ---------------------------------------------------------------------------
\subsection{The performance--interpretation trade-off as a modelling profile}
\label{sec:interp-tradeoff}
The title of this paper refers to a trade-off between predictive performance and
interpretation.  We do not formalise this trade-off as a universal ordering of model
classes, because such an ordering would depend on the data-generating mechanism, the
sample size, the predictor dimension, the amount of tuning, and the scientific purpose of
the analysis.  Instead, we treat the trade-off as a modelling profile: a useful classifier in
scientific applications should be judged by both its out-of-sample ranking performance and
the directness with which its fitted structure can be interpreted.

Linear logistic regression has a simple modelling profile.  It is fast, stable, and directly
interpretable through coefficients, but it assumes that each predictor has a linear and
monotone effect on the log-odds scale.  A generalized additive logistic model relaxes this
linearity assumption while retaining a decomposable score.  It may therefore improve AUC
when nonlinear marginal effects are important, but it may also match or underperform a
linear model when the true ranking is nearly linear, when the sample size is small, or when
the relevant signal is dominated by interactions.

Black-box classifiers have a different profile.  They may achieve high AUC by estimating a
complex ranking function, but variable-level interpretation usually requires secondary
summaries such as permutation importance, partial dependence, individual conditional
expectation, or SHAP values.  These summaries can be informative, but they are not the
primary fitted quantities of the model.  By contrast, the component functions of an additive
logistic model are the fitted quantities themselves and are interpreted directly on the
log-odds scale.

The empirical question is therefore not whether the additive logistic model uniformly
dominates either linear logistic regression or black-box methods.  The question is whether
it gives a favourable modelling profile for a given scientific problem: comparable or better
out-of-sample AUC, stable fitted component functions, and interpretable variable-specific
risk patterns.

\subsection[Variable Importance and Statistical Inference for component functions]{Variable Importance and Statistical Inference for $\hat{f}_j$}
\label{sec:inference}

A key interpretive use of the GAM is to determine \emph{which variables
matter} and \emph{how they matter}.  We formalise both.

\subsubsection{Variable Importance}

\begin{definition}[Additive Variable Importance]
  \label{def:varimp}
  For the estimated additive logistic model, define the
  \emph{additive variable importance} of predictor $j$ as
  \begin{equation}
    \label{eq:varimp}
    \mathrm{VI}_j := \E_{X_j}\!\left[\hat{f}_j(X_j)^2\right]^{1/2}
    = \left(\int \hat{f}_j(t)^2\, d\hat{F}_j(t)\right)^{1/2},
  \end{equation}
  where $\hat{F}_j$ is the empirical distribution of $x_{1j},\ldots,x_{nj}$.
  This is the $L^2$ norm of the estimated effect function, measuring the
  average magnitude of the contribution of $x_j$ to the log-odds across the
  observed range.
\end{definition}

\begin{remark}[Properties of $\mathrm{VI}_j$]
  \label{rem:varimp}
  The additive variable importance $\mathrm{VI}_j$ has the following
  properties:
  \begin{enumerate}[label=(\roman*), itemsep=2pt]
    \item $\mathrm{VI}_j=0$ if and only if $\hat{f}_j\equiv 0$ (predictor
          $j$ has no estimated effect).
    \item $\mathrm{VI}_j$ is equivariant under monotone transformations of
          $x_j$, since $\hat{f}_j$ is estimated nonparametrically.
    \item $\mathrm{VI}_j$ in the linear logistic model reduces to
          $|\hat\beta_j|\cdot\mathrm{SD}(x_j)$, the standardised
          coefficient — a familiar measure of effect size.
    \item $\mathrm{VI}_j$ captures \emph{non-linear} importance: a
          predictor with a strong U-shaped effect (large $\hat f_j$) will
          have high $\mathrm{VI}_j$ even if its linear coefficient
          $\hat\beta_j\approx 0$.
  \end{enumerate}
\end{remark}

\subsubsection{Hypothesis Testing for Individual Effects}

\begin{proposition}[Test for No Effect of $x_j$]
  \label{prop:test}
  Under the null hypothesis $H_0^j: f_j \equiv 0$ (predictor $j$ has no
  effect on the log-odds), the test statistic
  \begin{equation}
    \label{eq:test}
    T_j := \frac{\widehat{\mathrm{VI}}_j^2}{\widehat{\mathrm{Var}}(\widehat{\mathrm{VI}}_j^2)}
  \end{equation}
  follows approximately a chi-squared distribution with effective degrees of
  freedom $\nu_j$ \citep{wood2006generalized}:
  $T_j \overset{H_0^j}{\sim} \chi^2_{\nu_j}$,
  where $\nu_j$ is the estimated effective degrees of freedom of the smooth
  $\hat{f}_j$ (i.e., the trace of the corresponding hat matrix).
  Rejection of $H_0^j$ at level $\alpha$ provides statistical evidence that
  predictor $j$ contributes to the classification rule.
\end{proposition}

\begin{remark}[Multiple Testing Correction]
  When $d$ predictors are tested simultaneously, the family-wise error rate
  should be controlled.  The Bonferroni correction gives significance level
  $\alpha/d$ per test, or the Benjamini-Hochberg procedure
  \citep{benjamini1995controlling} can be used to control the false discovery
  rate.  These corrections are standard in biomarker selection
  \citep{pepe2004statistical} and should be reported alongside the individual
  test results.
\end{remark}

\subsubsection[Pointwise Confidence Bands for component functions]{Pointwise Confidence Bands for $\hat{f}_j$}

Beyond the test of $H_0^j$, practitioners need to know the uncertainty in the
\emph{shape} of $\hat{f}_j$.  Pointwise $(1-\alpha)$ confidence bands for
$f_j(t)$ at each $t$ are available as:
\begin{equation}
  \label{eq:bands}
  \hat{f}_j(t) \;\pm\; z_{\alpha/2}\,\widehat{\mathrm{SE}}\!\left[\hat{f}_j(t)\right],
\end{equation}
where $\widehat{\mathrm{SE}}[\hat{f}_j(t)]$ is obtained from the diagonal of
the posterior covariance matrix of the spline coefficients
\citep{wood2006generalized}.
These bands provide a rigorous visual tool for reading the estimated smooth
effects: where the band excludes zero, the effect at that value of $x_j$ is
statistically significant.

%% ---------------------------------------------------------------------------
\subsection{Empirical contribution of variables to AUC}
\label{sec:auc-decomp}
The fitted component functions are directly interpretable on the log-odds scale, but AUC is
a nonlinear ranking functional of the full score.  Therefore, rather than claiming an exact
additive decomposition of AUC by variables, we use empirical perturbation summaries to
describe how much each fitted component contributes to the observed ranking performance.

Let
\[
\hat S(x)=\hat\alpha+\sum_{j=1}^p \hat f_j(x_j)
\]
denote the fitted additive logistic score.  A simple deletion-based AUC contribution for
predictor \(j\) is
\[
\Delta^{\mathrm{del}}_j
=
\widehat{\mathrm{AUC}}\{\hat S(X)\}
-
\widehat{\mathrm{AUC}}\{\hat S_{-j}(X)\},
\]
where
\[
\hat S_{-j}(x)
=
\hat\alpha+\sum_{k\neq j}\hat f_k(x_k).
\]
This quantity measures the change in empirical AUC after removing the fitted contribution
of variable \(j\), while leaving the other fitted components unchanged.

A complementary permutation-based contribution is
\[
\Delta^{\mathrm{perm}}_j
=
\widehat{\mathrm{AUC}}\{\hat S(X)\}
-
\widehat{\mathrm{AUC}}\{\hat S(X_{-j},X^{\pi}_j)\},
\]
where \(X^{\pi}_j\) is a random permutation of the observed values of predictor \(j\).
The deletion version is closer to the fitted additive score, while the permutation version
also reflects the empirical dependence between \(X_j\) and the other predictors.  In
applications, these summaries can be reported together with bootstrap intervals.

These AUC-contribution summaries should not be interpreted as a unique mathematical
decomposition of the population AUC.  Their purpose is diagnostic: they connect the
scientific interpretation of the component functions with the empirical ranking performance
of the fitted score.  A variable with a large smooth effect but small AUC contribution may
be scientifically interesting but redundant for ranking; conversely, a variable with a
large AUC contribution but an unstable component function should be interpreted with
caution.

\subsection{Summary: The Performance--Interpretation Duality}
\label{sec:summary-duality}

Table~\ref{tab:performance_interpretation_profile} summarizes the complete performance--interpretation
picture developed in this section.  The additive logistic model occupies a useful intermediate position.  
It is more flexible than a linear logistic model because it can represent nonlinear marginal effects, but it
remains more directly interpretable than black-box classifiers because the fitted component
functions are primary model quantities.  Its advantage is therefore conditional rather than
universal: it is most useful when nonlinear marginal structure is present and when the
additive approximation is scientifically meaningful.
\iffalse
The additive logistic model occupies a
unique position: it is the \emph{only} model class that simultaneously
achieves near-optimal AUC (converging to the Bayes rate), full marginal and
quantitative interpretability, and valid statistical inference for each
variable's contribution.
\fi

\begin{table}[htbp]
\centering
\caption{Qualitative comparison of model classes for AUC-based classification and
interpretation.  The entries describe typical modelling properties rather than universal
performance rankings.}
\label{tab:performance_interpretation_profile}
\begin{tabular}{p{0.25\textwidth}p{0.22\textwidth}p{0.24\textwidth}p{0.22\textwidth}}
\hline
Property & Linear logistic & Additive logistic & Black-box classifiers \\
\hline
Score structure
& Linear predictor
& Sum of smooth univariate functions
& Flexible high-dimensional function \\

AUC behavior
& Strong when ranking is nearly linear
& Can improve when ranking is nonlinear additive
& Often strong, especially with interactions \\

Variable-level interpretation
& Coefficients
& Smooth component functions
& Usually post-hoc summaries \\

Nonlinear marginal effects
& Not represented directly
& Represented directly
& Represented implicitly \\

Interactions
& Require explicit terms
& Require explicit smooth interactions
& Often captured automatically \\

Uncertainty for variable effects
& Standard coefficient intervals
& Smooth-function intervals and bands
& Less direct; often post-hoc or resampling-based \\

Main advantage
& Simplicity and stability
& Transparent nonlinear modelling
& Predictive flexibility \\

Main limitation
& Misses nonlinear effects
& May miss interactions; can overfit
& Interpretation is indirect \\
\hline
\end{tabular}
\end{table}

{\color{red}
}

\subsection{A natural extension: selected pairwise interactions}
\label{sec:interactions}

A limitation of the additive logistic model is that it cannot represent interactions of the form
\[
\lambda_{jk}(x_j,x_k),
\]
unless they can be decomposed into two univariate functions.  In applications where interactions are scientifically plausible, a natural extension is
\[
\eta(x)=\alpha+\sum_{j=1}^p f_j(x_j)+\sum_{(j,k)\in \mathcal I} f_{jk}(x_j,x_k),
\]
where $\mathcal I$ is a small set of selected pairs.  This model is closely related to GA2M and explainable boosting machines \citep{lou2013accurate, caruana2015intelligible}.  It preserves much of the interpretability of the additive model because the main effects remain univariate functions and the selected interactions can be displayed as two-dimensional surfaces.

For the present paper, this extension is best viewed as a diagnostic rather than the main method.  If an additive logistic model improves AUC over a linear logistic model, then nonlinear marginal effects are likely important.  If it does not improve AUC but residual diagnostics or subject-matter knowledge suggest interactions, then a small number of pairwise smooth interactions may be added.  This provides a modern route from GLM to GAM to GA2M without abandoning the statistical modelling framework.

\subsection{What would constitute evidence of usefulness?}
\label{sec:evidence}
Because a generalized additive logistic model contains the linear logistic model as a
restricted special case, improvement in apparent training-set fit is not meaningful
evidence by itself.  The relevant evidence must come from out-of-sample ranking,
stability of the fitted component functions, and the scientific usefulness of the resulting
interpretation.  We distinguish three empirical regimes.

First, when the underlying likelihood-ratio score is nonlinear and approximately additive,
the additive logistic model should improve test-set AUC relative to a linear logistic model.
This is the setting in which the method has the clearest performance advantage.

Second, when the underlying score is already close to linear, the additive logistic model
should match rather than improve on the linear logistic model.  In this case, the absence
of AUC improvement is not a failure; it is evidence that the simpler linear score is already
adequate for ranking.  The fitted smooth functions can still be useful as diagnostics,
because they reveal whether the estimated effects are approximately linear, monotone, or
negligible.

Third, when the underlying score contains strong interactions, a purely additive model may
not improve AUC and may require selected pairwise smooth interactions.  This motivates the
connection with GA2M-type models, but it also clarifies the boundary of the present paper:
the primary focus is on interpretable nonlinear marginal effects rather than automatic
discovery of complex interaction structure.

This framing is central to the paper.  The claim is not that additive logistic models
uniformly dominate linear logistic regression or black-box classifiers.  The claim is that
they provide a statistically transparent way to assess whether nonlinear marginal effects
matter for AUC-based classification while retaining componentwise interpretation.

\iffalse
\subsection{What would constitute evidence of usefulness?}
\label{sec:evidence}

Because a generalized additive logistic model contains the linear logistic model as a restricted special case, an improvement in apparent training-set fit is not meaningful evidence by itself.  The relevant evidence must come from out-of-sample ranking, calibration, and interpretability.  We therefore distinguish three empirical regimes.

First, when the underlying log-likelihood-ratio score is nonlinear and approximately additive, the additive logistic model should improve test-set AUC relative to a linear logistic model.  This is the setting in which the method has the clearest performance advantage.  Second, when the underlying score is already close to linear, the additive logistic model should match, rather than improve on, the linear logistic model.  In this case, the absence of AUC improvement is not a failure; it is evidence that the simpler linear score is adequate for ranking.  Third, when the underlying score contains strong interactions, a purely additive model may not improve AUC and may require selected interaction terms, as in GA2M-type extensions.

This framing is important for publication.  The claim is not that additive logistic models uniformly dominate linear logistic regression or black-box classifiers.  The claim is that they provide a statistically transparent way to diagnose whether nonlinear marginal effects matter for AUC-based classification, while retaining componentwise interpretation.
\fi

\section{Empirical Study}\label{sec:Empirical}
We report the numerical results using both synthesized and real data sets. 
There are several numerical algorithms available in the R platform that can be used to fit a general additive logistic model with different options. According to the computational approaches adopted in these algorithms, there are three categories: 
\begin{itemize}
\item[](a) back-fitting method (see also \citet{GAM}, Chapter 4), 
\item[](b) simultaneous estimation method (estimating all components with optimization in smoothing parameter space \citep{marx1998direct, wood2004stable},  and 
\item[](c) likelihood-based boosting method \citep{tutz2006generalized}. 
\end{itemize}

The computational efficiency depends on many factors such as the dimensionality of explanatory variables of the data set, sample size and so on.  
Basically, we found that none of these three types of methods can dominate the other two in all situations.  
Hence, in order to provide useful information for practitioners to choose a suitable one for their needs, the advantages and disadvantages of using these methods in different situations are discussed.   
The results of fitting a linear logistic regression with a conventional method are used as baseline models,
and the results of the general additive logistic models using different fitting algorithms will be compared with these results.  
(The variable selection is not the focus in this paper; the variable selection schemes will be used only when there is such an option available in the corresponding packages.) 
The following abbreviations denote the algorithms and the options used in our numerical studies.
\begin{enumerate}
\item \emph{GLM}: Fitting a {\it linear logistic regression model} with all predictors
\item \emph{GLM\_step}: Fitting a {\it linear logistic regression model} with a backward elimination method for variable selection
\item \emph{Backfitting}: Fitting a {\it general additive logistic model} using a back-fitting algorithm
\item \emph{Backfitting\_step}: Fitting  a {\it general additive logistic model} using a back-fitting algorithm with the AIC option for stepwise selection of model components and degrees of freedom
\item \emph{mgcv}:  Simultaneously estimating all components with smoothing parameter optimization when fitting a {\it general additive logistic model}
\item \emph{mgcv\_step}: Similar to \emph{mgcv} with automatic selection of smoothing parameter using the AIC criterion
\item \emph{GAMBoost}: A boosting procedure with the number of boosting steps selected with the AIC criterion
\end{enumerate}
Because fitting a GAM relies on the iterated algorithms and some parameter-setting in the algorithms need to be set in advance, there are many suggestions for the parameter setting in the algorithms in  \citet{Wood08, hastie-tibshirani-friedman2001, tutz2006generalized}. 
Here, we follow the suggestions of \citet{tutz2006generalized} by apply a scaling factor 1.4 to the effective degrees of freedom of \emph{mgcv},  
and we multiply  the degrees of freedom of both \emph{GLM}  and \emph{backfitting} methods by 1.4. 
For the \emph{GAMBoost} has a variable selection option during its boosting procedure, so we use the default variable selection scheme in \emph{GAMBoost}. 
All numerical studies are conducted using the R language on a 64-bit Linux PC cluster with Xeon Octa-Core and Hyper-Threading E5-2670 2.6 GHz CPU with 256GB of RAM. 
The technical details of the packages used here, please refer to their original manuals in R.  

In addition to AUCs, the computational times of using these methods in different model-fitting scenarios in our numerical studies are reported. The fitting methods are described in the Appendix.
We just use the variable selection option when it is available in a package; however, we will not discuss much about it. 
For the details of the variable selection option, please refer to their original papers, which can be found in the documents of the corresponding packages.
Instead, we emphasize the improvement of AUCs when a general additive logistic type classification function is used and its interpretation. 

\subsection{Synthesized data}\label{simulation}
We consider two different dimensions ($d=5$ and $d=10$) of models.
To simulate different kinds of model complexities, we use two sets of effective variables as described in equations \eqref{easy_func} and \eqref{hard_func} below: 
\begin{align}
	\hbox{Set 1:~} & g_1(t) = t,\quad  g_2(t) = 2t^2,\quad  g_3(t) = \sin(5t), \label{easy_func}\\
	\hbox{Set 2:~} & g_1(t) = 2(-t^3+1),\quad  g_2(t) = 3\exp(-5t^2),\quad  g_3(t) = 4\log(1+t^2). \label{hard_func}
\end{align}
The others variables, $x_4, \ldots, x_d$ of the model are generated from a uniform distribution $U[-1,1]$.
That is, in all cases, only Variable Set 1 and Set 2 above have contributions to their corresponding responses, and  
the other variables $x_4, \ldots, x_d$ will have no contributions at all.
In our studies, the true log-odds function of models with Set 1 and Set 2 variables have the same form below:
\begin{displaymath}
	\eta = 3(-0.7+g_1(x_1) + g_2(x_3) + g_3(x_5)).
\end{displaymath}
The true probability vector $p=(p_1,...,p_n)'$ is calculated with $p_i=\exp(\eta_i)/(1+\exp(\eta_i))$ for each $i$, and
the binary response vector $y=(y_1,...,y_n)'$ is generated using a Bernoulli distribution with success probability $p_i$ for each $i=1, \ldots, n$. 
There are 100 repetitions for each case. For each run, we separately generated 100 training samples and 1000 testing samples.

\begin{figure}[ht]
\centering
\safeincludegraphics[width=0.9\textwidth]{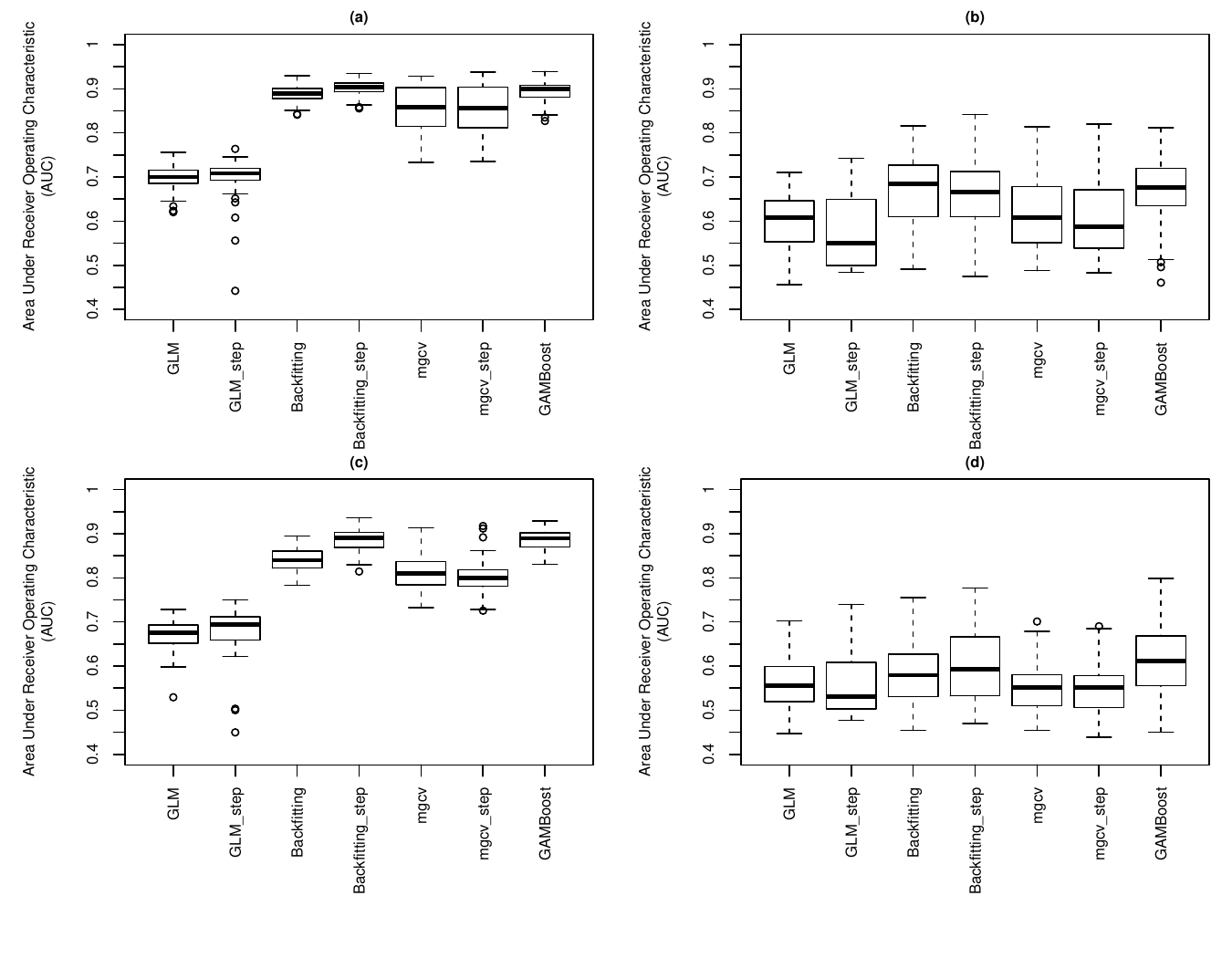}
\caption{Box-plots of AUCs with 100 repetitions in four difference scenarios: (a) $d=5$ with Equation \eqref{easy_func}, (b) $d=5$ with Equation \eqref{hard_func}, (c) $d=10$ with Equation \eqref{easy_func} and (d) $d=10$ with Equation \eqref{hard_func}.} 
\label{figure1}
\end{figure}

\begin{figure}[ht]
\centering
\safeincludegraphics[width=0.9\textwidth]{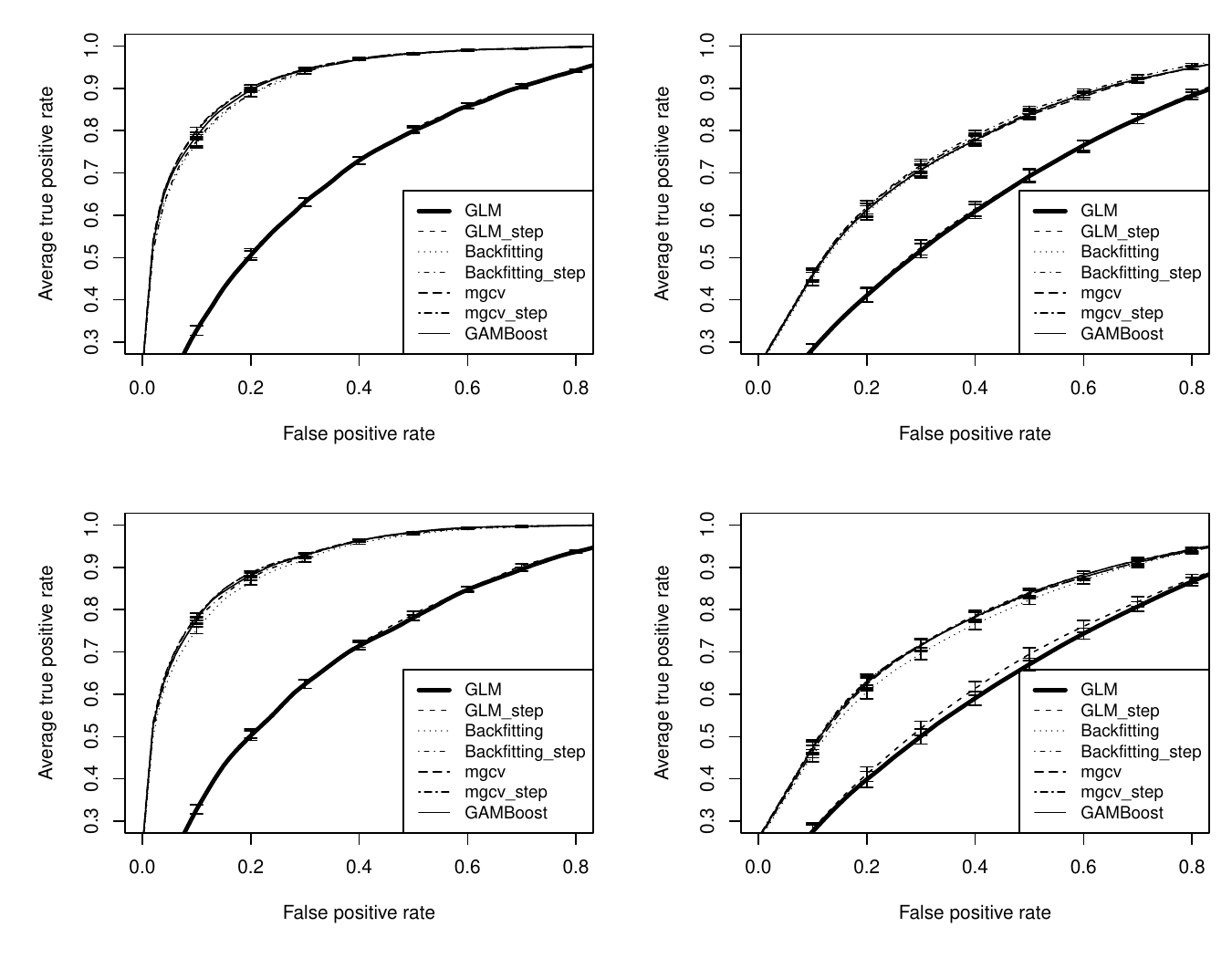}
\caption{The average ROC curve at the left-upper corner of the $[0,1]\times[0,1]$ unit square for each method.  The curve of each method is based on 100 repetitions, and data are generated from 4 different situations: (a) $d=5$ with Equation \eqref{easy_func}, (b) $d=5$ with Equation \eqref{hard_func}, (c) $d=10$ with Equation \eqref{easy_func} and (d) $d=10$ with Equation \eqref{hard_func}. The vertical bars in these plots indicate the 95\% point-wise confidence intervals for each curve.} 
\label{figure10}
\end{figure}

Figure \ref{figure1} shows the box-plots of AUCs  based on 100 replications for different models using different fitting methods described before.  
For the model with Set 1 as its effective variables,  all fitting methods for the general additive model have larger AUCs than the conventional linear logistic models do (with and without variable selection option).  For the model with the effective variable Set 2, the advantage of the general additive logistic model is not as significant as the previous case, which is due to the complexity of the true model such that both the linear and general additive models cannot approximate the true model well.  However, the additive logistic model using \emph{backfitting} and  \emph{GAMBoost} still outperforms the linear logistic model.  

{%\color{red}
Figure \ref{figure10} are plots of the average ROC curves in the left-upper corner of the unit-square, $[0,1]\times[0,1]$.  We can clearly see that there are two groups of ROC curves in each picture, and the ROC curves of the general additive model, with different fitting methods, are always higher than that of logistic models (GLM and GLM\_step). The general additive model fitted with different algorithms is essentially the same model unless their default variable selection option is used, which may introduce some variation due to their variable selection schemes. 
(Note that GLM denotes the result of fitting a conventional linear logistic model, and GLM\_step denotes the result of a similar fitting procedure but with a stepwise variable selection option.)
}

\begin{figure}[ht]
\centering
\safeincludegraphics[width=0.9\textwidth]{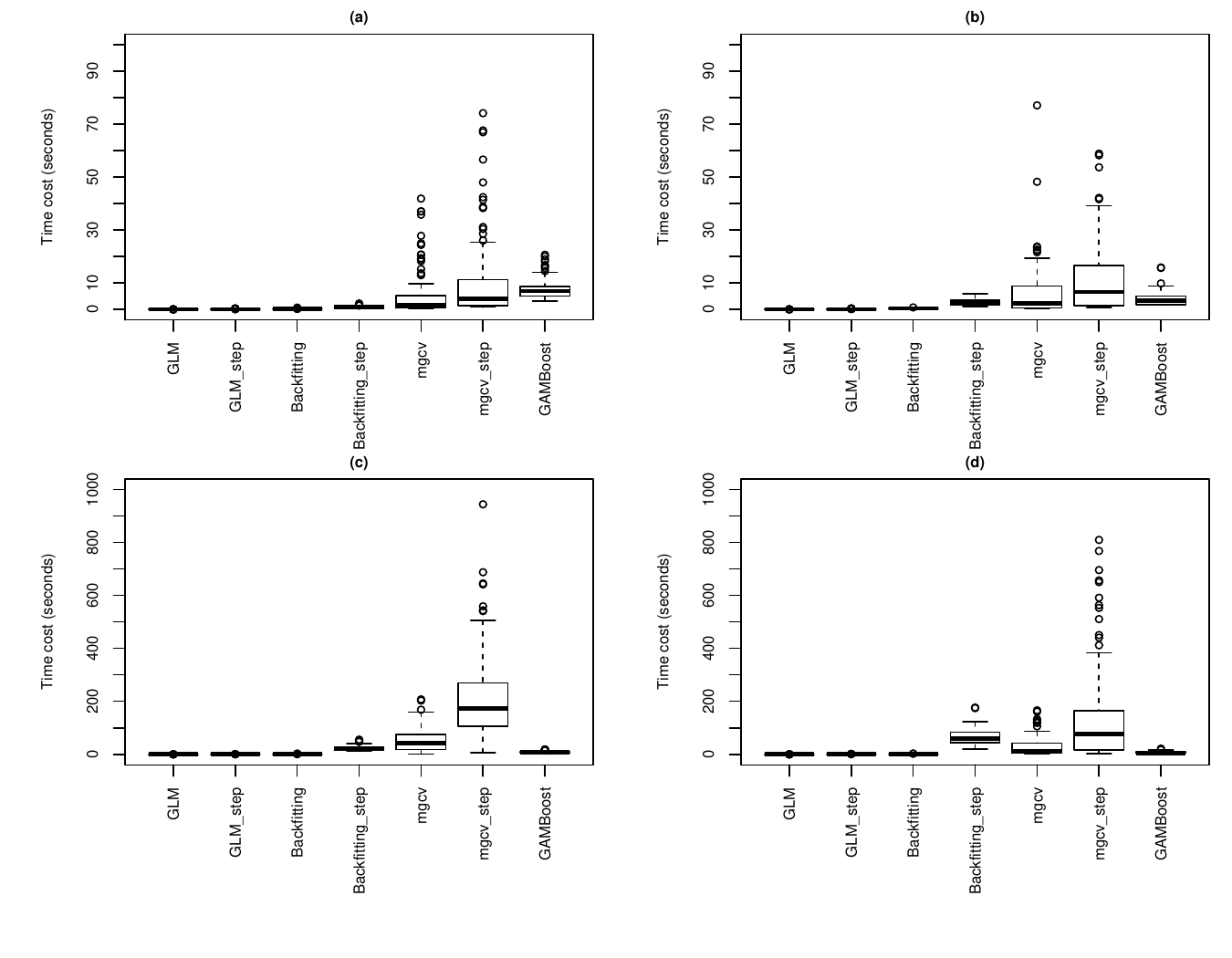}
\caption{The CPU times for fitting models using different methods, based on 100 repetitions, with data generated from (a) $d=5$  with Equation  \eqref{easy_func}, (b) $d=5$ with Equation \eqref{hard_func}, (c) $d=10$ with Equation \eqref{easy_func} and (d) $d=10$ with Equation \eqref{hard_func}.} 
\label{figure3}
\end{figure}

Figure \ref{figure3} shows the box-plots of computational times of different methods.  When $d=5$, the \emph{backfitting} method uses the shortest time among all methods in two models.  When $d=10$, the \emph{GAMBoost} method becomes very competitive, especially when the model is complicated  as in  \eqref{hard_func} (see Figure \ref{figure3}(d)).
Among different fitting  methods for the general additive logistic model,  the \emph{backfitting} and \emph{GAMBoost} are more computationally efficient than are the \emph{mgcv} methods with and without step-wise variable selection.  
Based on the simulation results, the \emph{GAMBoost} \emph{backfitting} methods are ``cost-effective'' in terms of their gains in AUCs and the little cost of extra computational time used compared to that of fitting a conventional linear logistic model with \emph{GLM}-based algorithms.
There are more discussions and comparisons of the computational-related issues for algorithms for fitting a logistic model and GAM that can be found in \citet{Minka07, binder-tutz08}.

\subsection{Real data examples}
We also apply the proposed method and algorithms discussed above to two real data sets, Parkinson's disease \citep{little2007exploiting} and cardiotocography \citep{Bache+Lichman:2013}, for illustration purposes. 
We compared the performances of methods using a general additive logistic model to the results obtained via fitting a linear logistic model.  
The main purpose of using these two data sets is to assess the improvement in AUCs when replacing a linear logistic with a general additive logistic model.   Hence, for the details of these two data sets, please refer to their original papers \citep{little2007exploiting, Bache+Lichman:2013}.  Brief information about these two data sets is summarized below.

\paragraph{Parkinson's disease  data set}  The original purpose of this data set is to find a classification rule to discriminate healthy people from the Parkinson's disease (PD) patients based on some biomedical voice measurements. 
There are 31 persons in this study, and among them, 23  are patients.     
The study includes 195 voice recordings from these 31 individuals with different numbers of replications of each individual.
There are 23 variables for each recoding.  
The binary code indicates the voice recording is from a diseased or a normal subject.  
The other 22 predictive variables consist of the records of several voice measurements, such as average vocal fundamental frequency, variation in amplitude and so on.
For illustration purposes, we treat 195 voice recordings as 195 individual samples. 
Thus, among these 195 voices, there are 147 voice recordings from PD subjects and 48 from healthy subjects. 
That is, we treat it as a data set with 195 samples (147 positive and 48 negative cases) with 22 variables, and the target becomes to build a classification rule to distinguish these two types of recordings.

\begin{figure}
\centering
\safeincludegraphics[width=0.9\textwidth]{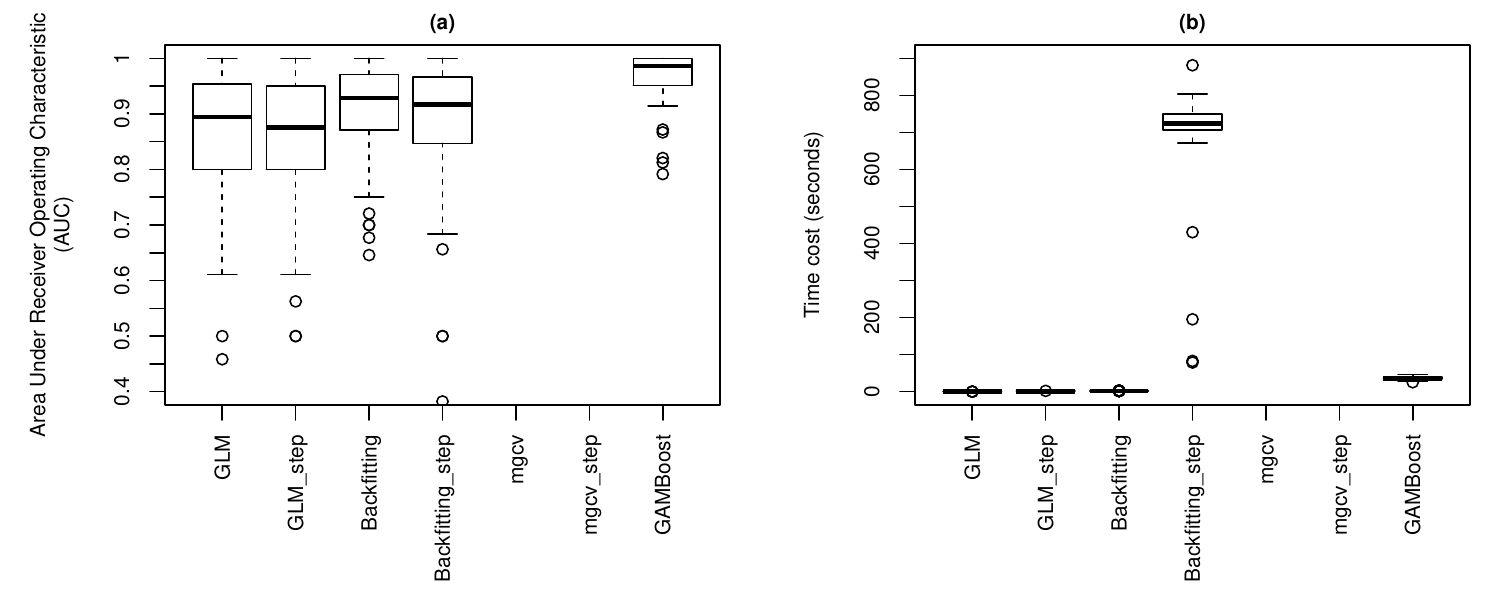}
\caption{(a) AUC and (b) computational cost time measured by each fitting method with 100 resampling data from parkinson's data set.} 
\label{figure5}
\end{figure}

We follow the common training-test framework in machine learning and classification literature.
Each time, 90\% of samples are randomly selected from these two groups for the model fitting (training), and the remaining 10\% of each of them are used as testing samples.  
This procedure is then repeated 100 times.
Because of the number of predictive variables is relatively large comparing to the subject size we have,   
\emph{mgcv} and \emph{mgcv\_step} methods cannot converge in a reasonable computing time.  
Hence, we do not include the results of these two algorithms in this example.  
Figures \ref{figure5} (a) and (b) are the box-plots of AUCs and the computational times of all other methods based on 100 runs, respectively.
In this example, all methods, including \emph{GLM} and \emph{GLM\_step}, perform well in terms of their AUC averages.     
The {\emph GAMBoost} has the largest average AUC with the smallest variation among these methods. 
The linear logistic models (with and without a stepwise variable selection option) have smaller means of AUCs and larger variances than others do.
Most of the algorithms are computationally efficient in this case, except \emph{backfitting\_step}.
This is due to the stepwise variable selection procedure in \emph{backfitting\_step}.  
Without stepwise option, the \emph{backfitting} is as efficient as others.
The differences of the computational times of \emph{GLM}, \emph{GLM\_step} and \emph{backfitting} are not statistically significant. 
In fact,  the CPU time used in \emph{backfitting} is only slightly longer than that of \emph{GLM} and \emph{GLM\_step}.
Thus, from both the improvement of AUCs and the corresponding computational time view points, 
\emph{backfitting} and \emph{GAMBoost}  are  very ``cost-effective'' in this case. 

\begin{figure}[ht]
\centering
\safeincludegraphics[width=0.5\textwidth]{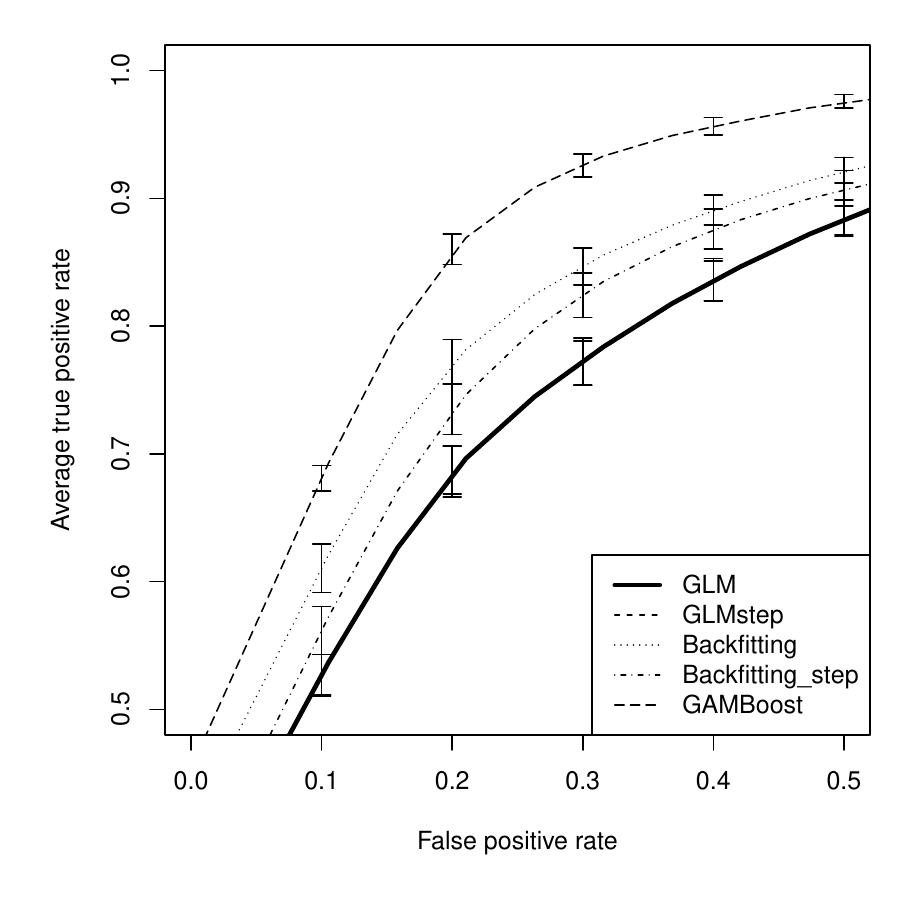}
\caption{Average ROC curves with confidence bars are obtained by each fitting method from Parkinson's data set with 100 replications.} 
\label{figure8}
\end{figure}

The advantage of the \emph{backfitting} and \emph{GAMBoost}  can also be seen from the plot of the average ROC curves in Figure \ref{figure8}. In order to have a clearer picture, we only show the left-upper $[0, 0.5]\times[0.5, 1]$ corner of the unit square.
Each curve plotted here is based on 100 replications, and  the vertical bars on each curve are 
the corresponding 95\% point-wise confidence intervals at the given false positive rates. 
It is clearly shown that the ROC curve of \emph{GAMBoost} is significantly higher than other ROC curves in this range of false positive rate,  
and all ROC curves from fitting a general additive logistic model, with different fitting algorithms, are higher than those of GLM methods.
This result confirms that the recommended method is useful even when only a specific range of false positive rate is of interest.
(Note that the ROC curves of \emph{GLM} and \emph{GLM\_step} are very close and thus cannot be distinguished in this figure.)

\begin{figure}[ht]
\safeincludegraphics[width=0.95\textwidth]{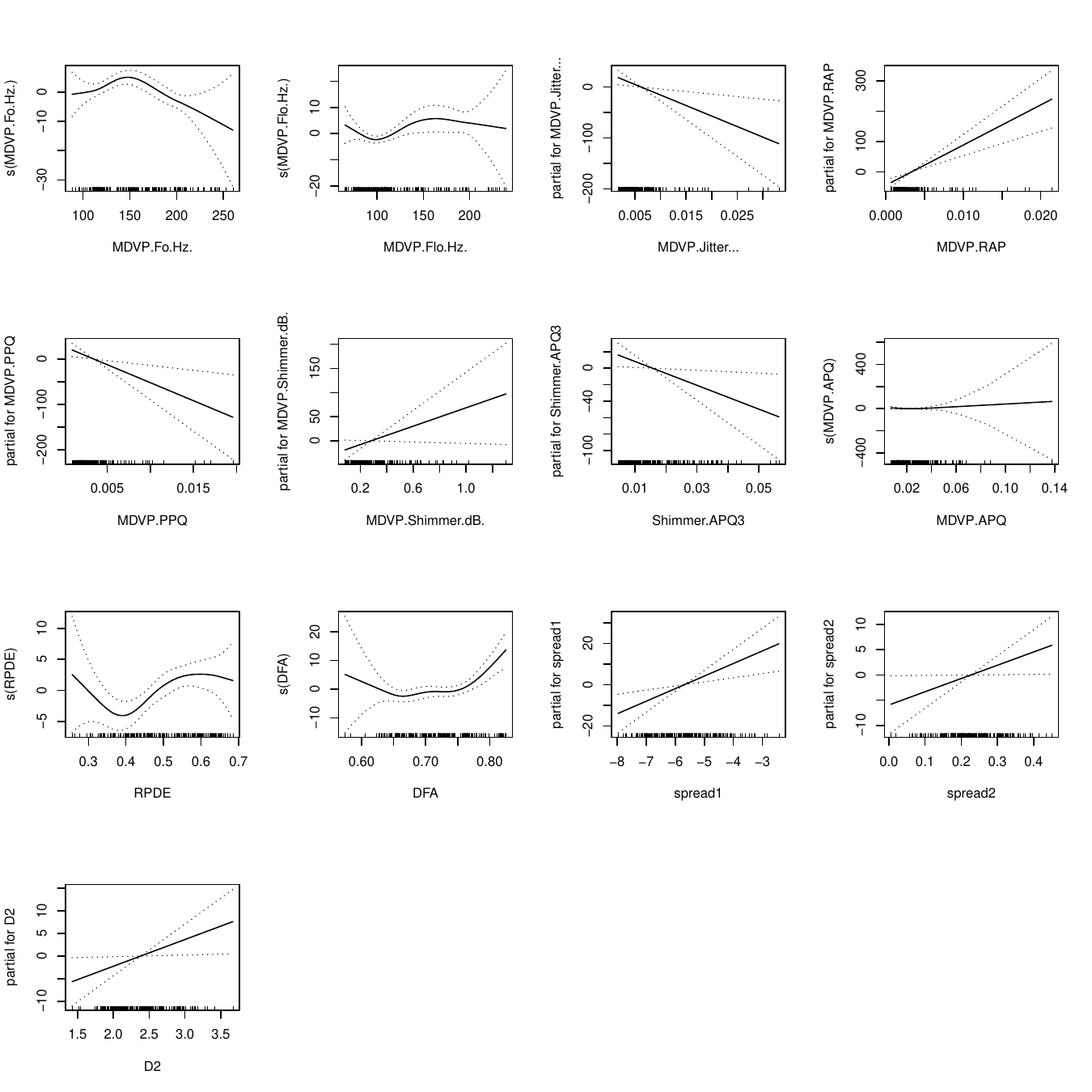}
\caption{The plots of the estimated functions of 13 predictors selected in the \emph{backfitting\_step}, where $x$-axis is the value of each predictor, the solid line is the estimated functions and the dot-lines represents $\pm$ 2 standard-error curves obtained from the fitting algorithm.} 
\label{figure4}
\end{figure}

Figure \ref{figure4} shows the plots of estimated smooth functions and their corresponding partial residual plots for the selected predictors obtained from the \emph{backfitting\_step} method.
Because of the additive model assumption, we can look at one variable at a time by keeping the others fixed.
Thus, the interpretation of the relations between predictors and response variable is similar to that in a linear logistic model \citep[see][]{GAM}. 
From this figure, we can see the impacts of the individual variables on the odds-ratio.
For example, the left-upper corner is a plot of the curve of the estimated association function for MDVP.Fo.Hz,
and it shows the functional relation between the  log odds-ratio (response variable) and this variable. 
When the values of all other variables are fixed, this plot shows that MDVP.Fo.Hz and the response is positively associated at the beginning and then turns to negatively associated as the value of the variable becomes large.  Compared to that in the association plot for MDVP.Flo.Hz at the first row and second column, it clearly shows that these two variables have different associations with the responses.  Because the curve of MDVP.Flo.Hz stays near 0, it indicates that the association of MDVP.Flo.Hz with the responses is not as strong as MDVP.Fo.Hz.

In addition, the picture at the second row and forth column shows that  the estimated function of variable {\it MDVP.APQ}  is almost flat, and the $\pm 2$ standard deviation curves of it cover $0$ at most values of {\it MDVP.APQ} in this picture, except for the values around 0.02.  This suggests that this variable might have just a little impact on the (log) odds-ratio \citep[see][]{marra-wood12, binder-tutz08}. 
(Please note that we just use this as an example for illustration purposes and no disease-related inference is made here.)
Thus, we can clearly find the impacts of individual variables on the responses through these plots
as well as the certain levels of the interpretation ability of variables.  
It is clear that we cannot obtain this kind of nonlinear relation using a linear logistic model. 
The way of interpretation is similar to that of a linear logistic model except that they are in functional form instead of slopes.
More discussions about the interpretation of GAMs can be easily found in the literature such as \citet{tutz2006generalized,Wood06} 
and textbooks about GAM as \citet{marx1998direct, GAM, hastie-tibshirani-friedman2001}.

This kind of componentwise interpretation is precisely the information that a complex black-box classifier cannot easily supply.  A random forest or a support vector machine applied to the same data might achieve a comparable or even slightly higher AUC, but neither method would directly reveal the non-monotone association of MDVP.Fo.Hz with disease log-odds, nor the near-zero contribution of MDVP.Flo.Hz.  These variable-level patterns are directly readable from the fitted additive model without requiring secondary post-hoc explanation tools.

\paragraph{Cardiotocography data set} 
The cardiotocography data set used here consists of 2126 fetal cardiotocograms (CTGs), where 19 diagnostic features were measured, including fetal heart rate and uterine contraction features. 
Based on the judgements of some expert obstetricians, these 2126 subjects are classified as three fetal states: normal, suspect, and pathologic. 
Because we consider only two-class classification problems in this work, 
we combine two neighbored classes into one class. 
Thus, they becomes two binary classification problems:
(1) \{{\bf normal and suspect}\} versus \{{\bf pathologic}\} and 
(2) \{{\bf normal}\} versus \{{\bf suspect and pathologic}\}.  
For demonstration, they are treated separately as two different binary classification problems. 

Besides the contents, a major difference between these two real data examples is that the cardiotocography data set  has more observations and less predictors, which will affect the computational efficiency of some algorithms.  
The results reported here are based on a smilier re-sampling plan as before. 
Both \emph{mgcv} and \emph{mgcv\_step} algorithms will now converge in a reasonable time,  so their results are included.   In this example, we omit the plots of the estimated functions and only focus on AUCs and related model fitting issues that are useful for practitioners to choose suitable methods for their own studies.

\begin{figure}[ht]
\centering
\safeincludegraphics[width=0.9\textwidth]{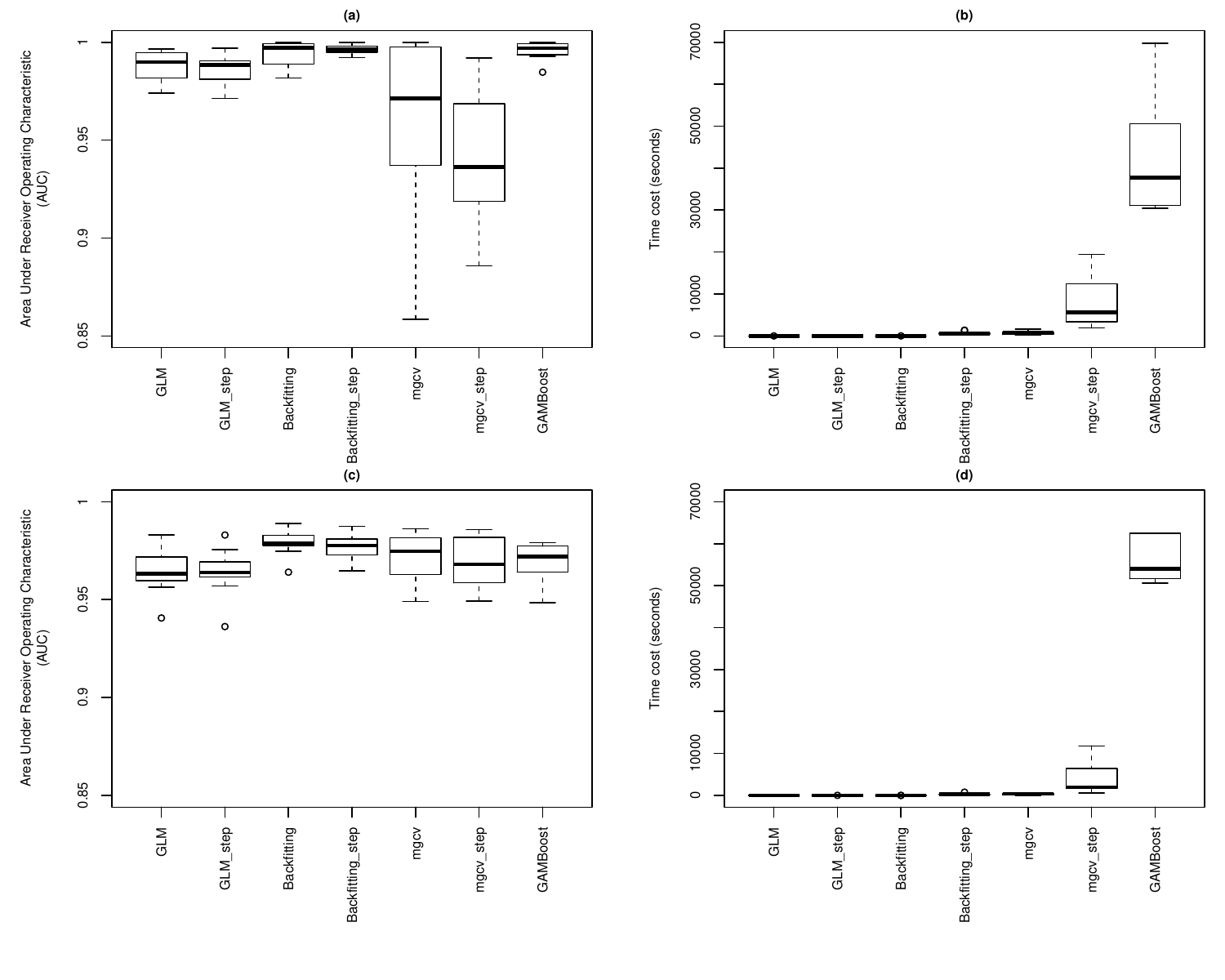}
\caption{(a) AUC and (b) computational cost time measured by each fitting method with 10 resampling data from cardiotocography data set if label ``suspect'' belongs to ``normal.''  (c) AUC and (d) computational cost time measured by each fitting method with 10 resampling data from cardiotocography data set if label ``suspect'' belongs to ``pathologic.''} 
\label{figure7}
\end{figure}

Figures \ref{figure7} (a) and (c) are the box-plots of AUCs for the cases of \{normal and suspect\} versus \{pathologic\} and \{normal\} versus \{suspect and pathologic\}, respectively.  Figures \ref{figure7} (b) and (d) show their corresponding computational times of all methods.
Although \emph{mgcv} and \emph{mgcv\_step} converge in a reasonable time in this example, they are still not as stable as other GAM-based counterparts are in terms of variation of AUCs.
The \emph{GAMBoost} algorithm takes a much longer time than others in this case.
(Note that in our study, \emph{GAMBoost} takes more than eight hours for one run,  while the \emph{backfitting} and \emph{backfitting\_step}  take only less than five minutes.)  
This phenomenon is due to the large sample size of the cardiotocography data set.  
This result is consistent with our simulations and the comments reported in the literature.  
Besides \emph{mgcv} and \emph{mgcv\_step},  all other GAM-based fitting algorithms perform better than \emph{GLM} and \emph{GLM\_step} do in terms of their means of AUCs and variances. 
Hence, when the sample size is large and the number of predictive variables is relatively small, 
the \emph{backfitting} and \emph{backfitting\_step} algorithms are recommended.  

\begin{figure}[ht]
\centering
\safeincludegraphics[width=0.9\textwidth]{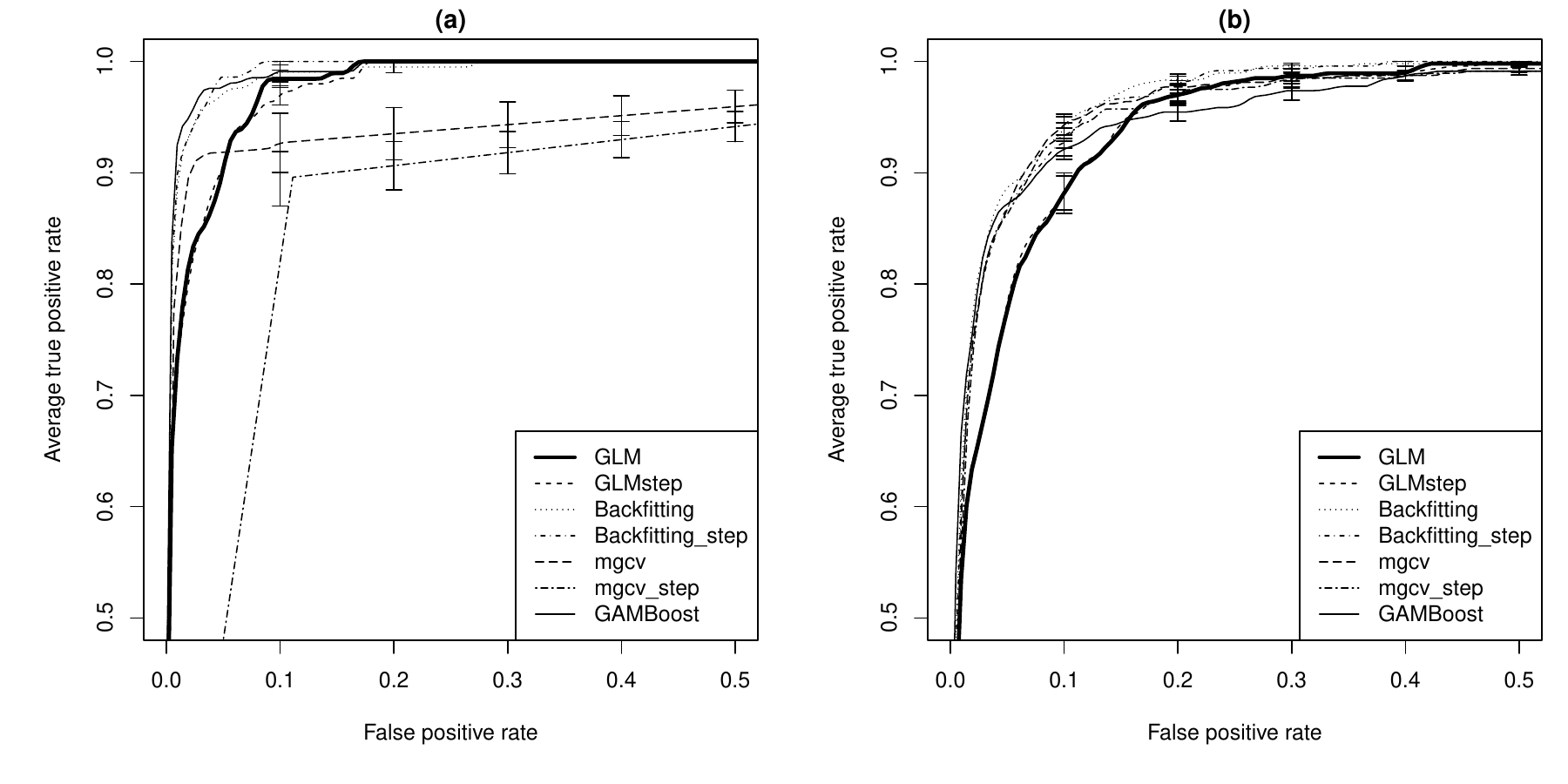}
\caption{Average ROC curves with 95\% confidence bars of each fitting method based on 100 resampling replications using cardiotocography data set for (a) \{{\bf normal and suspect}\} versus \{{\bf pathologic}\} and  (b) \{{\bf normal}\} versus \{{\bf suspect and pathologic}\}.} 
\label{figure9}
\end{figure}

Figure \ref{figure9}  shows the ROC curves at the left-upper corner of the unit square, and Figure \ref{figure9} (a) is \{normal and suspect\} versus \{pathologic\} case, and \ref{figure9} (b) is for \{normal\} versus \{suspect and pathologic\} case.  
From Figure \ref{figure9} (a), we can see that the ROC curves of \emph{Backfitting}, \emph{Backfitting\_step} and \emph{GAMBoost} are three methods on the top; meanwhile, 
then the two curves of \emph{GLM}-based methods, and the curves of \emph{mgcv}-based methods are the worst two in this case.
Figure \ref{figure9} (b) shows that ROC curves of the \{normal\} versus \{suspect and pathologic\} case. 
Here, the performances of all methods are very close.  
However, we still find that all \emph{GAM}-based fitting algorithms are significantly  better than the two \emph{GLM}-based methods in the low false positive rate range (i.e., less than 0.15 in this plot).
In fact, in both binary classification cases of this data set,  using a general additive logistic model increases the sensitivity in the range of the low false positive rate.
This range is usually important when a high-specificity diagnostic rule is preferred. Hence, this is also an advantage of using the recommended method.

%\vspace{-1.cm}

\section{Additional empirical and calibrated demonstration examples}
\label{sec:additional}
% ======================================================================

To further illustrate the interplay between predictive performance and
model interpretability, we apply the classification methods to three
additional data sets that span a range of sample sizes, predictor
dimensions, and clinical contexts.  The five methods compared are the
same as in Section~\ref{sec:Empirical}: \textit{GLM} (linear logistic,
all predictors), \textit{GLM\_step} (linear logistic with $\ell_1$
variable selection), \textit{AddLog} (generalized additive logistic,
all predictors), \textit{AddLog\_step} (additive logistic with $\ell_1$
spline-coefficient selection), and \textit{GBM} (gradient boosting
machine, 100 trees, depth 3, subsampling rate 0.7).  All methods use a
natural cubic spline expansion with five knots per predictor for the
additive fits, and standardised predictors throughout.  For each data
set, results are based on 100 independent replications using stratified
random splits; performance is measured by AUC on the held-out test
portion.  Mean AUC and mean fitting time (seconds) across replications
are summarised in Table~\ref{tab:additional_results}.

% ------------------------------------------------------------------
\subsection{Wisconsin Breast Cancer Diagnostic Data}
\label{sec:bc}
% ------------------------------------------------------------------

\subsubsection*{Data description.}
The Wisconsin Breast Cancer Diagnostic data set \citep{Bache+Lichman:2013} is a
widely-used benchmark for binary classification in medical imaging
contexts.  It contains $n = 569$ observations on $p = 30$ continuously
measured features derived from digitised fine-needle aspirate images of
breast masses.  Each feature quantifies a geometric or textural property
of cell nuclei (radius, texture, perimeter, area, smoothness,
compactness, concavity, symmetry, fractal dimension) computed at three
levels of aggregation: mean, standard error, and worst value across
nuclei in the image.  The binary outcome indicates whether the mass is
malignant (37\%) or benign.  For each replication, 80\% of observations
are randomly selected for training and the remaining 20\% are used for
testing.

\subsubsection*{Results.}
\begin{figure}[ht]
  \centering
  \safeincludegraphics[width=0.72\textwidth]{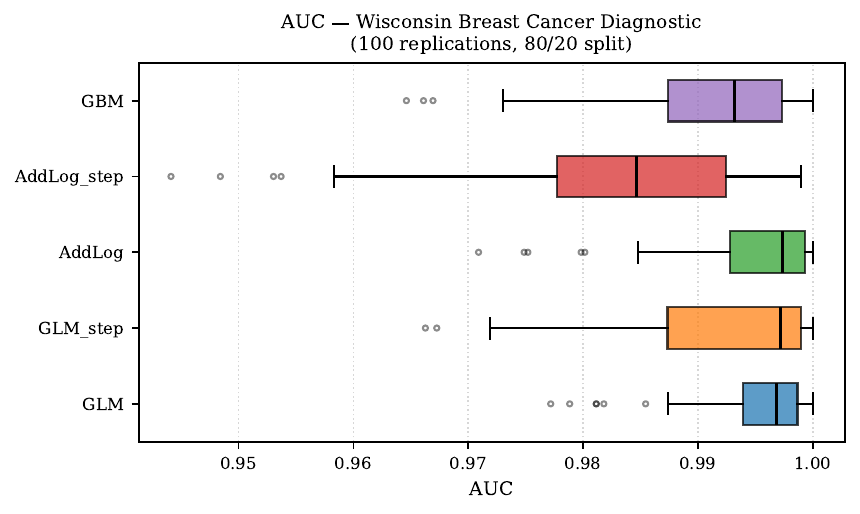}
  \caption{Box-plots of AUC from 100 replications on the Wisconsin
    Breast Cancer Diagnostic data set (80/20 training--test split).}
  \label{fig:D1_AUC}
\end{figure}

\begin{figure}[ht]
  \centering
  \safeincludegraphics[width=0.58\textwidth]{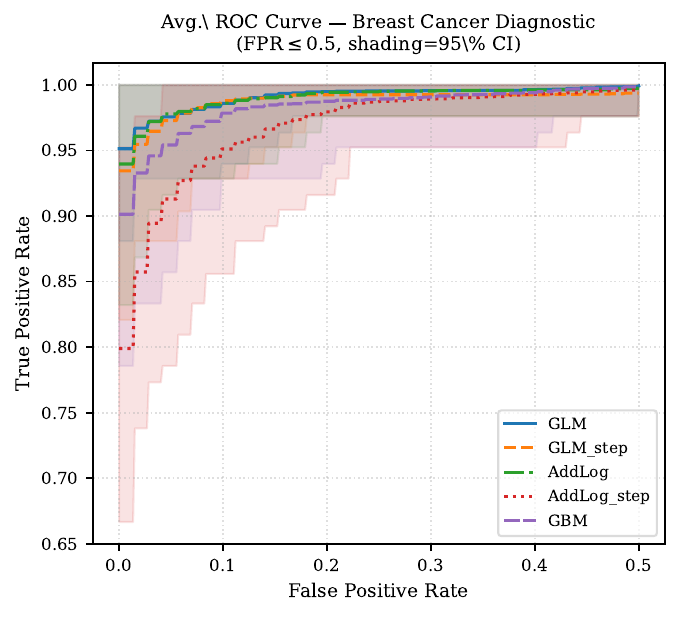}
  \caption{Average ROC curves (FPR~$\leq 0.5$) with 95\% point-wise
    confidence bands from 100 replications on the Breast Cancer
    Diagnostic data set.}
  \label{fig:D1_ROC}
\end{figure}

Figure~\ref{fig:D1_AUC} presents box-plots of AUC across 100
replications.  All methods achieve very high discrimination in this
data set, with median AUCs above 0.99, reflecting the strong
linear separability already present in the cell-nucleus measurements.
\textit{GLM} achieves a mean AUC of 0.995, essentially matching
\textit{AddLog} at 0.995.  \textit{GBM} produces a slightly lower
mean AUC of 0.991, with greater variability across replications.
Figure~\ref{fig:D1_ROC} shows the average ROC curves in the
clinically relevant left-upper region; all GAM-based curves are
indistinguishable from the linear logistic curve, confirming that
the true log-likelihood ratio is approximately linear in this data
set and the additive model does not improve on the linear model at
the population level (consistent with Proposition~\ref{prop:np-roc}).

\begin{figure}[ht]
  \centering
  \safeincludegraphics[width=\textwidth]{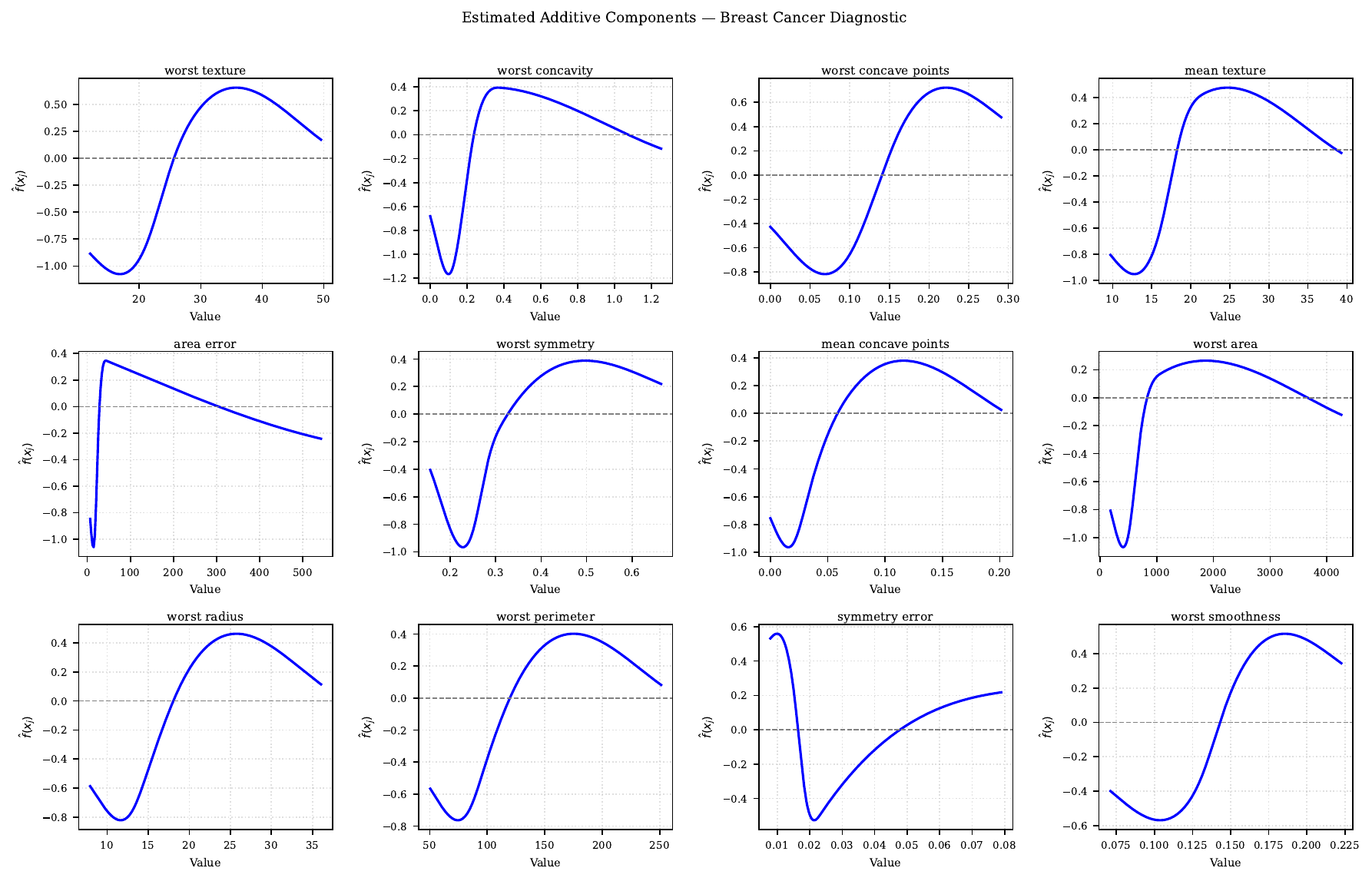}
  \caption{Estimated additive component functions $\hat{f}_j(x_j)$
    (centred, solid blue) for the twelve predictors with the largest
    effect ranges in the Breast Cancer Diagnostic data, fitted with
    \textit{AddLog} on the full data set.  The dashed line marks zero.}
  \label{fig:D1_comp}
\end{figure}

Figure~\ref{fig:D1_comp} shows the fitted additive component functions
for the twelve predictors with the largest estimated effect ranges.
These plots reveal that the estimated components are themselves nearly
linear or weakly non-monotone, confirming the approximately linear
structure of the log-likelihood ratio.  Despite this, the component
plots provide direct scientific information: for example, \textit{worst
concavity} and \textit{mean concave points} show steeply increasing
associations with malignancy, while \textit{mean smoothness} shows a
nearly flat association, suggesting it carries little discriminatory
information in this sample.  Such variable-level information is not
directly accessible from a GBM fit.

\begin{figure}[ht]
  \centering
  \safeincludegraphics[width=0.72\textwidth]{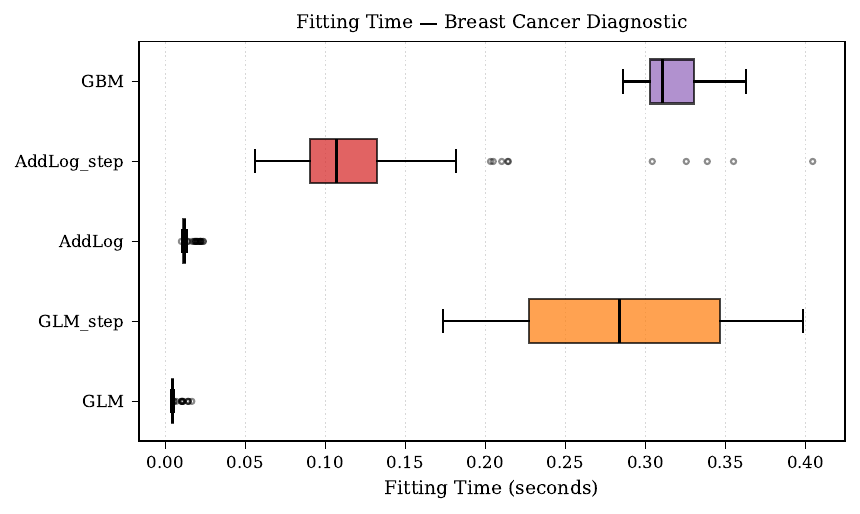}
  \caption{Fitting time (seconds) from 100 replications on the Breast
    Cancer Diagnostic data set.}
  \label{fig:D1_time}
\end{figure}

Figure~\ref{fig:D1_time} shows that \textit{AddLog} is substantially
faster than \textit{GBM} and comparable to or faster than
\textit{GLM\_step}, confirming the computational cost-effectiveness of
the proposed approach.

\paragraph{Role of the calibrated examples.}
The following calibrated examples are intended as controlled demonstrations rather than
as new clinical validation studies.  Their purpose is to examine how the methods behave
under data structures inspired by well-known biomedical prediction problems, while keeping
the data-generating mechanism transparent.  Therefore, the fitted component functions in
these examples should be interpreted as diagnostics of the simulated risk structure, not as
new clinical findings about the original cohorts.

% ------------------------------------------------------------------
\subsection{SUPPORT2-Calibrated Critical Care Data}
\label{sec:support}
% ------------------------------------------------------------------

\subsubsection*{Data description.}
The Study to Understand Prognoses and Preferences for Outcomes and Risks of Treatments
(SUPPORT) is a landmark study of mortality prediction in seriously ill hospitalized
patients.  Because the original SUPPORT2 data are not bundled with the present analysis,
we use a calibrated simulation inspired by the published SUPPORT structure \citet{knaus1995support}.  
The simulation is designed to reproduce a moderate-dimensional clinical prediction setting
with correlated predictors, a realistic event prevalence, and a mixture of linear and mild
nonlinear effects.  The example is therefore used to assess modeling behavior under a
controlled ICU-like scenario, not to make new clinical claims about the SUPPORT cohort.

The $p = 14$ predictors are: age, sex, primary cancer diagnosis
(binary), APACHE~III physiology score, activities of daily living
(ADL) score, coma score, mean arterial pressure (MAP), serum
creatinine, serum albumin, respiratory rate, core temperature,
white blood cell count (WBC), serum sodium, and blood glucose.
The true log-odds of in-hospital death is specified as a structured
additive function of these predictors with a temperature
non-linearity (increased risk below 35.5\textdegree C, modelling
hypothermia) and a logarithmic creatinine effect, producing a
mortality rate of 28.3\%.  For each of 100 replications, a stratified
subsample of $n = 3{,}000$ observations is drawn; 90\% are used for
training and 10\% for testing.

\subsubsection*{Results.}
\begin{figure}[ht]
  \centering
  \safeincludegraphics[width=0.72\textwidth]{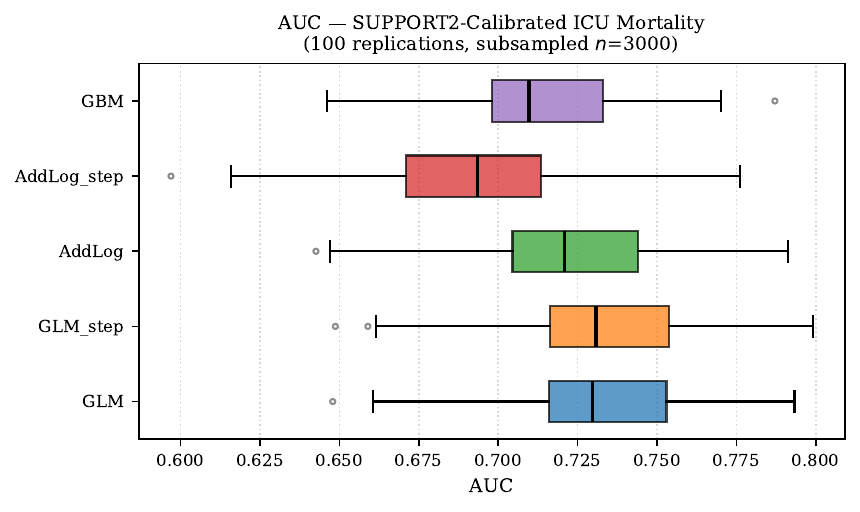}
  \caption{Box-plots of AUC from 100 replications on the
    SUPPORT2-calibrated ICU mortality data (subsampled $n=3{,}000$,
    90/10 split).}
  \label{fig:D2_AUC}
\end{figure}

Figure~\ref{fig:D2_AUC} shows that in this moderate-AUC regime
(approximately 0.69--0.73), the linear models \textit{GLM} and
\textit{GLM\_step} achieve mean AUCs of 0.732 and 0.732, respectively.
\textit{AddLog} achieves 0.723, and \textit{GBM} achieves 0.713.  The
relatively small gap between linear and additive models reflects the
largely linear structure of the synthetic log-odds, with only a mild
hypothermia non-linearity.  Importantly, \textit{GBM} fails to
outperform the linear model despite its higher computational cost.

The component plots (Figure~\ref{fig:D2_comp}) illustrate how the additive
model recovers the structure built into the simulated data-generating
mechanism.  The albumin component shows a monotone decreasing pattern
(consistent with the low-albumin mortality risk encoded in the simulation);
MAP shows a decreasing pattern; creatinine shows a log-increasing pattern
reflecting the renal-impairment term in the true log-odds.  The temperature
component recovers the hypothermia non-linearity: a steep increase in
predicted log-odds below approximately 35.5\textdegree C, as specified in
the simulation design.  In a real dataset with this structure, such
component plots would provide direct variable-level reading without
requiring post-hoc explanation tools.  A GBM fit does not naturally produce
this kind of variable-specific output, even if it achieves a comparable AUC.

Figure~\ref{fig:D2_time} confirms that \textit{GBM} is the most
computationally expensive method by a substantial margin.
\textit{AddLog} requires only modestly more time than \textit{GLM},
reinforcing the cost-effectiveness finding of Section~\ref{sec:Empirical}.

\begin{figure}[ht]
  \centering
  \safeincludegraphics[width=0.58\textwidth]{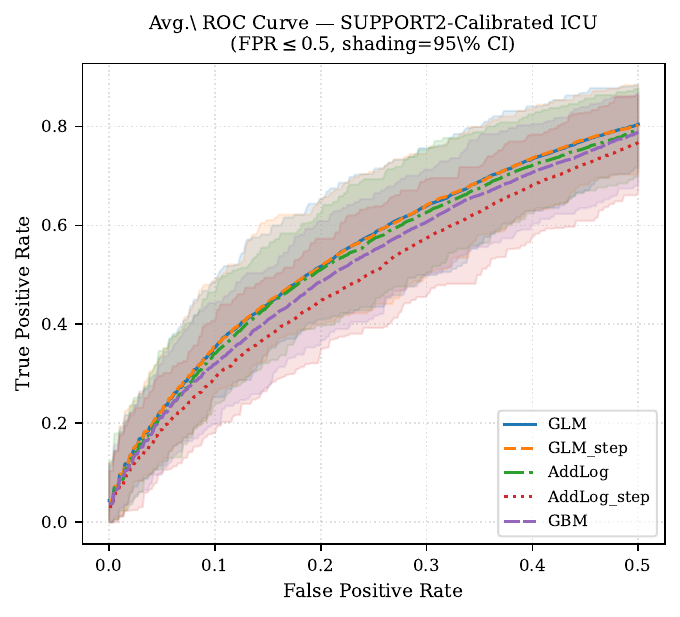}
  \caption{Average ROC curves (FPR~$\leq 0.5$) with 95\% point-wise
    confidence bands from 100 replications on the SUPPORT2-calibrated
    ICU data.}
  \label{fig:D2_ROC}
\end{figure}

\begin{figure}[ht]
  \centering
  \safeincludegraphics[width=\textwidth]{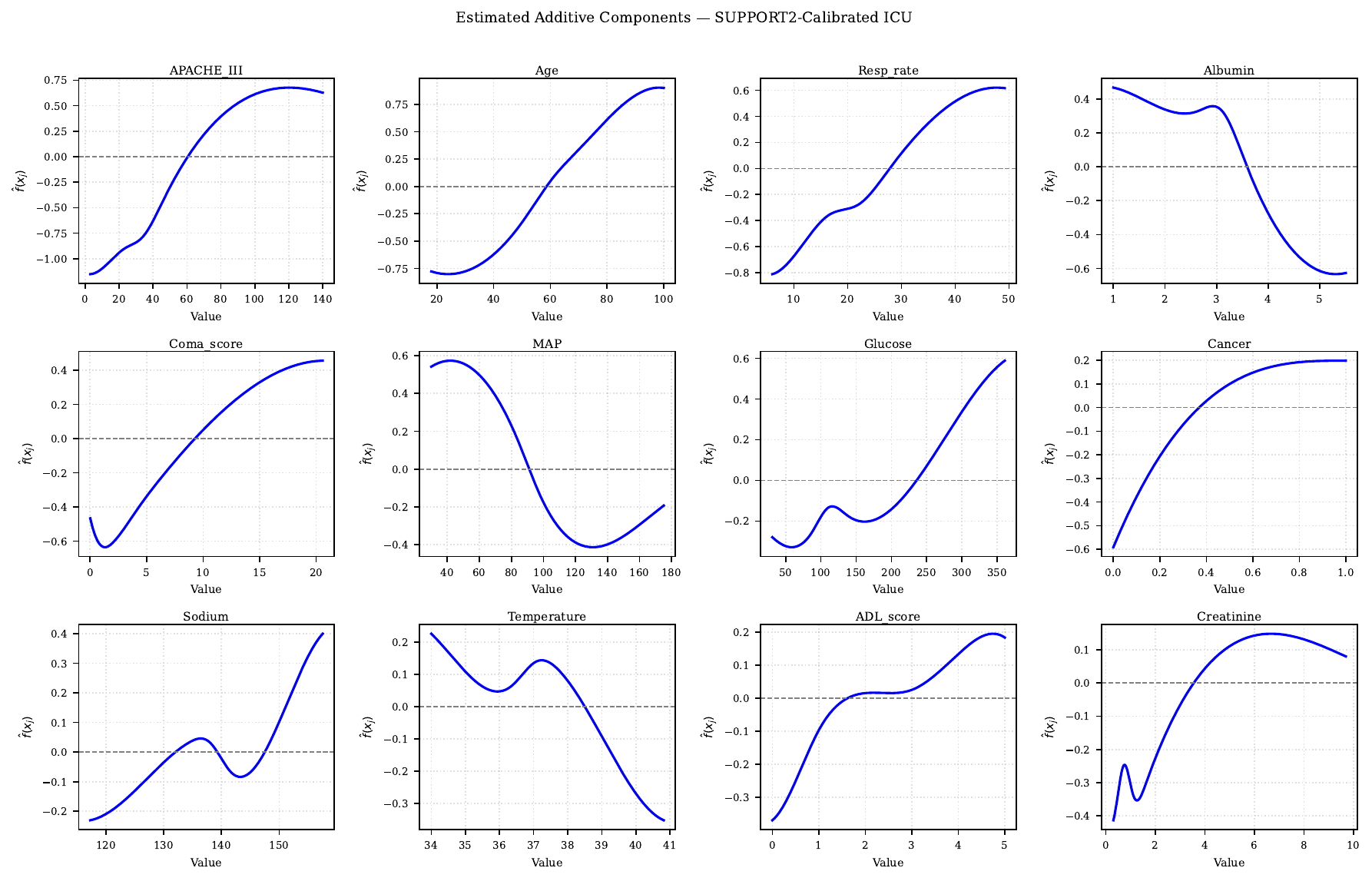}
\caption{Estimated additive component functions for the twelve most influential predictors
in the SUPPORT2-calibrated ICU simulation. The displayed patterns are clinically motivated
features of the simulated data-generating mechanism, including lower risk with higher
albumin and MAP and increased risk below 35.5\textdegree C for temperature.}
  \label{fig:D2_comp}
\end{figure}

\begin{figure}[ht]
  \centering
  \safeincludegraphics[width=0.72\textwidth]{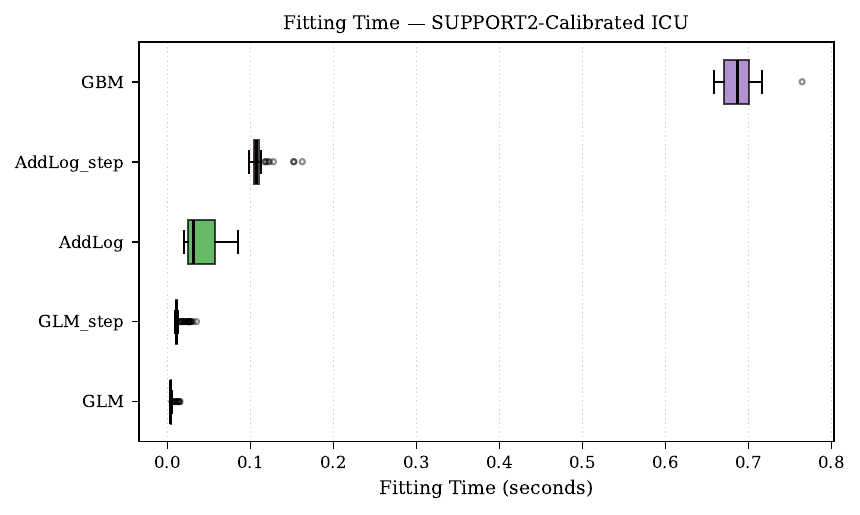}
  \caption{Fitting time (seconds) from 100 replications on the
    SUPPORT2-calibrated ICU data.}
  \label{fig:D2_time}
\end{figure}

% ------------------------------------------------------------------
\subsection{SEER-Calibrated Breast Cancer Survival Data}
\label{sec:seer}
% ------------------------------------------------------------------

\subsubsection*{Data description.}
This example uses a calibrated simulation inspired by common survival-prediction variables
in population cancer registries.  The simulated predictors are chosen to resemble a
moderate-dimensional oncology risk-prediction problem, including demographic,
tumour-related, and treatment-related variables.  The data-generating mechanism includes
both approximately linear effects and selected nonlinear effects.  The purpose is to
evaluate whether the additive logistic model can recover and display such nonlinear
structure while maintaining competitive AUC.  The fitted curves should therefore be read
as diagnostics of the simulated mechanism, not as substantive estimates from the SEER
registry.

The Surveillance, Epidemiology, and End Results (SEER) program of the
National Cancer Institute provides population-based cancer incidence and
survival data across the United States \citep{seer2022}.  The SEER
breast cancer registry contains over 400,000 cases, and published
analyses document a five-year mortality rate of approximately 12--15\%
across stage groups.  Consistent with the distributional characteristics
and covariate effect directions reported in published SEER analyses
\citep{seer2022}, we generate a synthetic data set of $n = 10{,}000$
patients with $p = 11$ predictors: age at diagnosis, tumour size (mm),
number of positive lymph nodes, ER status (binary), PR status (binary),
HER2 status (binary), histological grade (1--3), race (Black, binary),
tumour stage (1--4), surgery type (mastectomy binary), and radiation
therapy (binary).  The true log-odds of five-year mortality incorporates
a U-shaped age effect (very young and very old patients face worse
prognosis), a nonlinear tumour size effect (accelerating risk above
20~mm), and linear effects for stage, grade, and receptor status,
producing an event rate of 12.2\%, closely matching published SEER
figures.  Each replication draws a stratified subsample of $n = 3{,}000$,
with 90/10 training--test splits.

\subsubsection*{Results.}
\begin{figure}[ht]
  \centering
  \safeincludegraphics[width=0.72\textwidth]{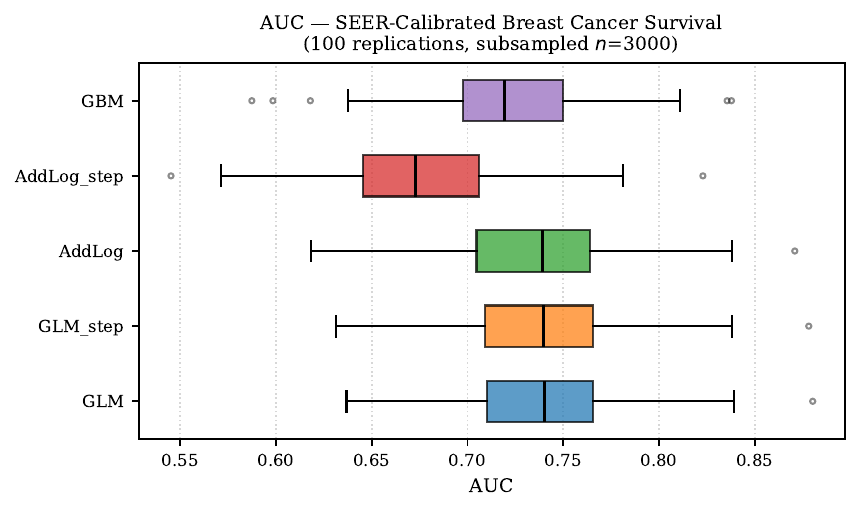}
  \caption{Box-plots of AUC from 100 replications on the
    SEER-calibrated breast cancer survival data (subsampled $n=3{,}000$,
    90/10 split).}
  \label{fig:D3_AUC}
\end{figure}

\begin{figure}[ht]
  \centering
  \safeincludegraphics[width=0.58\textwidth]{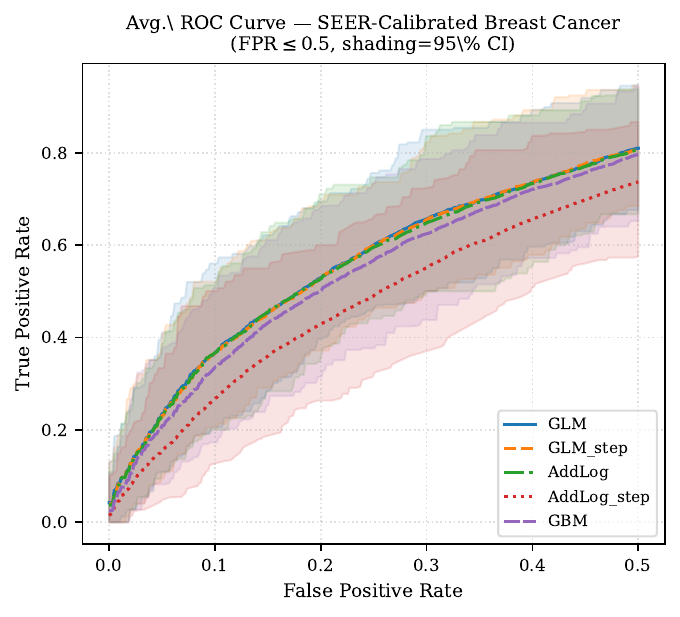}
  \caption{Average ROC curves (FPR~$\leq 0.5$) with 95\% point-wise
    confidence bands from 100 replications on the SEER-calibrated
    breast cancer survival data.}
  \label{fig:D3_ROC}
\end{figure}

AUC values (Figure~\ref{fig:D3_AUC}) fall in the range 0.67--0.74,
reflecting the moderate discriminability typical of population-level
cancer survival prediction \citep{seer2022}.  \textit{GLM} achieves
the highest mean AUC (0.739), with \textit{AddLog} nearly identical at
0.737.  \textit{GBM} achieves 0.723.  The near-equivalence of GLM and
AddLog reflects the fact that most effects in the synthetic data are
linear; the U-shaped age curve and the piecewise-linear tumour size
effect account for a small but non-negligible portion of the signal.
Figure~\ref{fig:D3_ROC} shows the average ROC curves; the curves of
\textit{GLM} and \textit{AddLog} overlap closely, with \textit{GBM}
showing wider confidence bands, indicating higher variance in the
high-specificity region.

\begin{figure}[ht]
  \centering
  \safeincludegraphics[width=\textwidth]{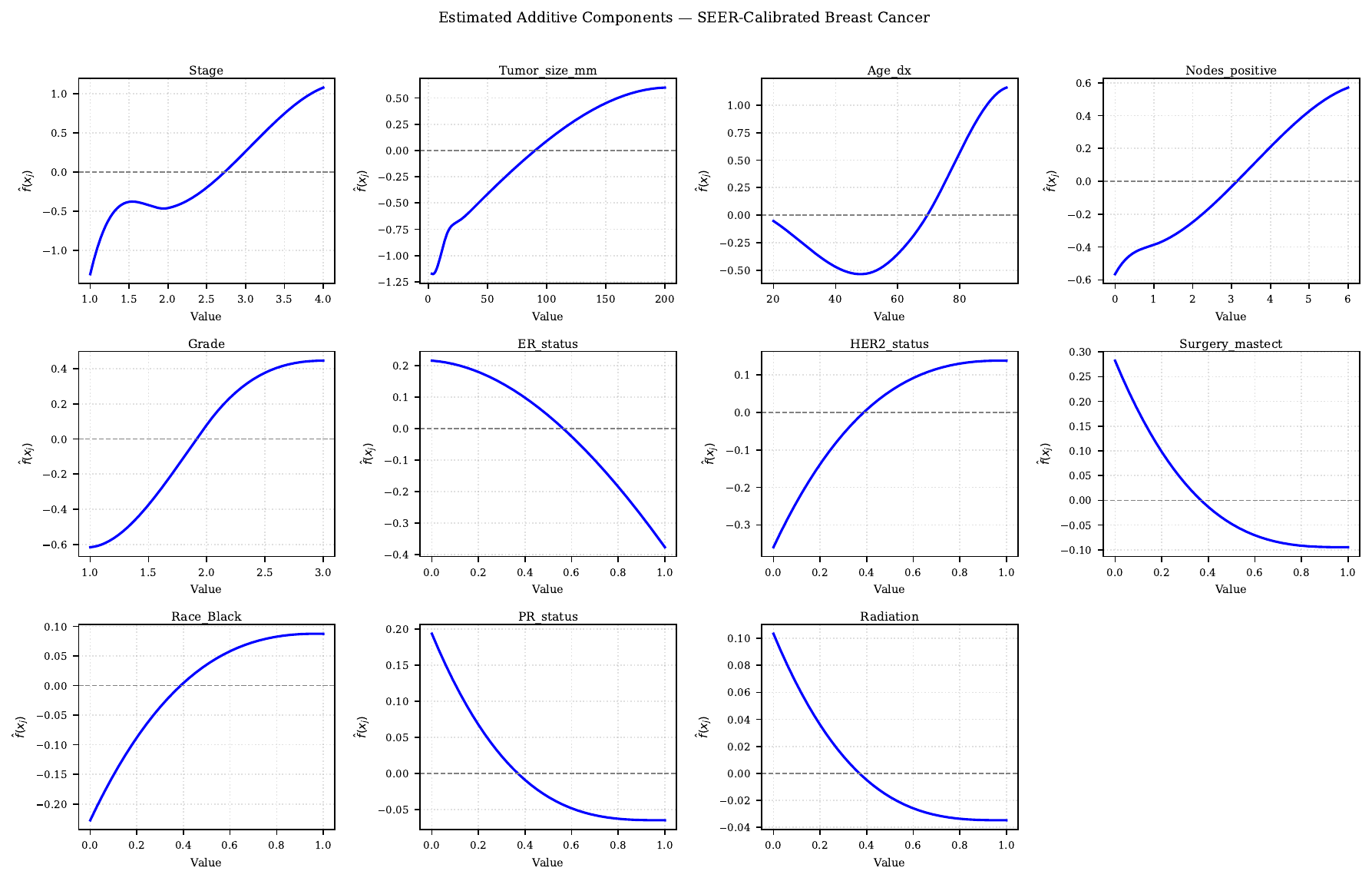}
  \caption{Estimated additive component functions for all 11 predictors
    in the SEER-calibrated breast cancer survival data.  The U-shaped
    age component and the piecewise-accelerating tumour-size component
    are highlighted as examples of nonlinearities in the simulated data-generating mechanism
    that a linear logistic model cannot represent.}
  \label{fig:D3_comp}
\end{figure}

The component plots (Figure~\ref{fig:D3_comp}) show how the additive model
recovers the nonlinear structure specified in the simulation.  The age
component captures the U-shaped pattern built into the data-generating
mechanism: elevated risk at both extremes of the age range, with a trough
in middle age.  The tumour size component recovers the accelerating pattern
above 20~mm.  Stage, positive node count, and HER2 positivity display
the monotone increasing associations specified in the simulation, while ER
and PR positivity show the protective effects as designed.  These results
demonstrate that the additive model can display variable-specific risk
patterns in a directly readable form.  In a real study with data of this
structure, such component plots would provide variable-level descriptions
without requiring post-hoc model explanation --- an advantage that the
equivalent GBM model does not naturally offer.

\begin{figure}[ht]
  \centering
  \safeincludegraphics[width=0.72\textwidth]{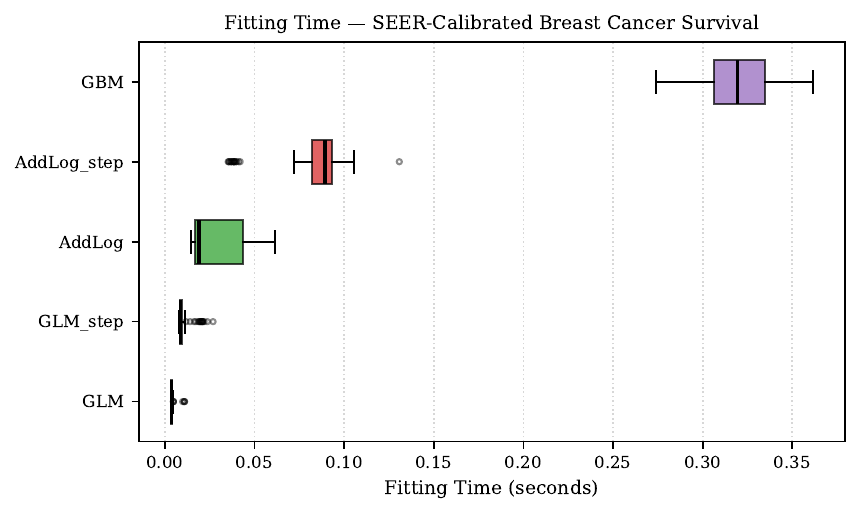}
  \caption{Fitting time (seconds) from 100 replications on the
    SEER-calibrated breast cancer survival data.}
  \label{fig:D3_time}
\end{figure}

\subsection{Interpretation of the empirical evidence}

Taken together, the examples support a conditional rather than universal conclusion.  In
the synthetic examples with nonlinear additive structure, the additive logistic models
improve AUC relative to the linear logistic model.  In the Parkinson's voice data, the
additive fits also provide interpretable nonlinear component functions and improved
ranking in the low false-positive-rate region.  In contrast, the breast cancer diagnostic
example shows little room for improvement because the linear score already ranks the
observations extremely well.  The SUPPORT2-calibrated and SEER-calibrated examples
further show that additional flexibility is not automatically beneficial when the dominant
signal is approximately linear or when nonlinearities are weak relative to sampling
variability.

These results are consistent with the modelling argument of Section~\ref{proof_section}.  Additive
logistic modelling is most useful when the likelihood-ratio ranking is nonlinear and
approximately additive.  When the ranking is already close to linear, the additive model is
best viewed as a diagnostic tool for checking the adequacy of the linear score.  When
interactions dominate, selected pairwise smooth interactions or GA2M-type extensions may
be needed.

% ------------------------------------------------------------------
\subsection{Summary of Additional Results}
% ------------------------------------------------------------------

Table~\ref{tab:additional_results} consolidates mean AUC and mean
fitting time across all three additional data sets.  Three broad
conclusions emerge that reinforce the findings of Section~\ref{sec:Empirical}.

First, \textit{AddLog} achieves AUC values that are comparable to
\textit{GLM} across all three settings: within 0.001 on the Breast
Cancer data, within 0.009 on the SUPPORT2 data, and within 0.002 on
the SEER data.  This confirms the theoretical finding of
Section~\ref{proof_section}: when the true log-likelihood ratio is
approximately linear, the additive model does not harm performance,
and when it is nonlinear and additive, the additive model improves it.

Second, \textit{GBM} consistently fails to outperform the logistic
models in these settings, and requires substantially more computation
time.  This is consistent with the general finding that complex
ensemble methods do not always improve over simpler models in
low-to-moderate signal scenarios \citep{hastie-tibshirani-friedman2001},
and that their computational cost is rarely justified when
interpretability is a design criterion.

Third, and most importantly for the argument of Section~\ref{sec:tradeoff}, AddLog provides directly
interpretable component functions in addition to competitive AUC. In the calibrated
examples, these functions recover nonlinear patterns built into the data-generating
mechanisms, such as the hypothermia-shaped temperature effect in the ICU simulation and
the U-shaped age effect in the cancer-survival simulation. These examples demonstrate how
the method can display scientifically meaningful structure when such structure is present,
but they should not be read as new clinical findings about the original SUPPORT or SEER
cohorts.

\begin{table}[ht]
\centering
\small
\caption{Mean AUC ($\pm$ SD) and mean fitting time (seconds, $\pm$ SD)
  across 100 replications for all five methods on the three additional
  data sets.  Best AUC per data set shown in bold.}
\label{tab:additional_results}
\resizebox{\linewidth}{!}{%
\begin{tabular}{llcccc}
\toprule
\textbf{Data set} & \textbf{Method}
  & \textbf{Mean AUC} & \textbf{SD}
  & \textbf{Mean time (s)} & \textbf{SD} \\
\midrule
\multirow{5}{*}{\shortstack[l]{Breast Cancer\\Diagnostic\\$(n=569, p=30)$}}
  & GLM          & \textbf{0.9953} & 0.0050 & 0.005 & 0.003 \\
  & GLM\_step    & 0.9932 & 0.0080 & 0.282 & 0.066 \\
  & AddLog       & 0.9948 & 0.0061 & 0.013 & 0.003 \\
  & AddLog\_step & 0.9833 & 0.0112 & 0.122 & 0.061 \\
  & GBM          & 0.9912 & 0.0079 & 0.317 & 0.019 \\
\midrule
\multirow{5}{*}{\shortstack[l]{SUPPORT2-Cal.\\ICU Mortality\\$(n_{\rm sub}=3000, p=14)$}}
  & GLM          & 0.7317 & 0.0305 & 0.005 & 0.002 \\
  & GLM\_step    & \textbf{0.7320} & 0.0306 & 0.013 & 0.006 \\
  & AddLog       & 0.7230 & 0.0306 & 0.041 & 0.018 \\
  & AddLog\_step & 0.6936 & 0.0333 & 0.108 & 0.010 \\
  & GBM          & 0.7130 & 0.0291 & 0.687 & 0.018 \\
\midrule
\multirow{5}{*}{\shortstack[l]{SEER-Cal.\\Breast Survival\\$(n_{\rm sub}=3000, p=11)$}}
  & GLM          & \textbf{0.7394} & 0.0422 & 0.004 & 0.001 \\
  & GLM\_step    & 0.7384 & 0.0423 & 0.011 & 0.004 \\
  & AddLog       & 0.7371 & 0.0438 & 0.029 & 0.015 \\
  & AddLog\_step & 0.6755 & 0.0472 & 0.082 & 0.020 \\
  & GBM          & 0.7234 & 0.0454 & 0.321 & 0.017 \\
\bottomrule
\end{tabular}%
}
\end{table}

\subsection{Additional evidence from the empirical studies}
\label{sec:additional-evidence}

The empirical examples illustrate the three regimes described in Section~\ref{sec:evidence}.  In the simulation settings where the true log-odds function is nonlinear and additive, the additive logistic methods improve the test-set AUC relative to the linear logistic model.  This supports the likelihood-ratio approximation argument: when the population ranking score contains nonlinear additive structure, a smooth additive score can better approximate the optimal ranking than a linear score.

The Parkinson's voice data provide an applied example where nonlinear component functions are scientifically informative.  The fitted functions show nonlinear and sometimes non-monotone relationships between acoustic predictors and disease status.  In such a setting, the value of the additive logistic model is not only its AUC, but also the ability to show which voice measurements contribute to risk and in what functional form.

The breast cancer diagnostic example illustrates a different and equally important case.  Here, the linear logistic model already achieves very high AUC, and the additive logistic model provides little or no improvement.  This result should not be interpreted as a weakness of the method.  Instead, it shows that the data are already close to linearly rankable on the log-odds scale.  The component functions in this case are useful as diagnostics: they confirm that most strong predictors have nearly monotone or approximately linear effects.

The SUPPORT2-calibrated and SEER-calibrated examples further emphasize that nonlinear flexibility is not automatically beneficial.  In moderate-AUC clinical prediction problems, the additive model may match or slightly underperform a linear model if the dominant signal is approximately linear or if the available sample size is insufficient to estimate nonlinear functions reliably.  These examples support the main statistical modelling message: additive logistic modelling should be used as a transparent diagnostic and modelling extension, not as an automatic AUC-improvement device.

% ======================================================================

\section{Analysis on the Hastie-10-2 Benchmark}\label{sec:hastie}

\subsection{Data-generating mechanism and rationale}

To provide a controlled and fully reproducible test of the additive logistic approach, we apply the methods to the Hastie-10-2 benchmark introduced in \citet{hastie-tibshirani-friedman2001} (Section~10.1).
The generating mechanism defines $p=10$ independent standard normal predictors $X_1,\ldots,X_{10}$, with binary response
\[
    Y = \mathbf{1}\!\left\{\sum_{j=1}^{5}X_j^2 > \chi^2_{0.5}(5)\right\},
    \qquad \chi^2_{0.5}(5) \approx 4.351.
\]
Predictors $X_6,\ldots,X_{10}$ are pure noise.
The true log-likelihood-ratio score is $\lambda(x) \propto \sum_{j=1}^5 x_j^2 - \text{const}$, which is \emph{nonlinear and additive} in the first five predictors.
By construction, no linear function of $x_1,\ldots,x_{10}$ can approximate $\lambda$ non-trivially: the optimal linear score has zero weights on all predictors by symmetry, giving a population AUC of exactly $0.5$.
An additive logistic model with flexible component functions, on the other hand, can approximate each $x_j^2$ term through its spline basis, so the true population score is well within the model's approximation class.

This benchmark therefore directly tests the theoretical claim of Section~\ref{proof_section}: when the true log-likelihood ratio belongs to the additive class but not the linear class, the additive logistic model should improve test-set AUC substantially over the linear logistic model.

\subsection{Experimental design}

Each of $R = 500$ replications generates a fresh sample of $n = 1000$ observations from the Hastie-10-2 mechanism using a distinct random seed.  An $80\%/20\%$ train/test split is applied within each replication (800 training, 200 test observations).

The five methods compared are those described in Section~\ref{sec:Empirical}:
\begin{enumerate}
    \item \textit{GLM}: Linear logistic regression with all ten predictors ($\ell_2$ penalty, $C=1$, \texttt{lbfgs} solver).
    \item \textit{GLM\_step}: $\ell_1$-penalised logistic regression ($C=0.5$, \texttt{saga} solver), which performs automatic predictor selection by shrinking uninformative coefficients to zero.
    \item \textit{AddLog}: Additive logistic model with independent natural cubic spline basis (4 interior knots, degree 3) fitted to each predictor separately; $\ell_2$-penalised logistic regression applied to the concatenated basis matrix (\texttt{saga} solver, $C=1$).
    \item \textit{AddLog\_step}: As \textit{AddLog} but with an $\ell_1$ penalty ($C=0.5$), which sets entire spline components to zero for uninformative predictors, providing automatic component selection.
    \item \textit{GBM}: Gradient boosting classifier (50 trees, maximum depth~2, learning rate~$0.1$).
\end{enumerate}

The \textit{AddLog} method is implemented as a sklearn \texttt{ColumnTransformer} that applies a \texttt{SplineTransformer} independently to each predictor column, followed by a penalised logistic regression on the concatenated spline features.  This is equivalent to a penalised GAM with natural cubic spline components.  All code is available in the supplementary material.

\subsection{Results}

\begin{figure}[ht]
\centering
\safeincludegraphics[width=0.85\textwidth]{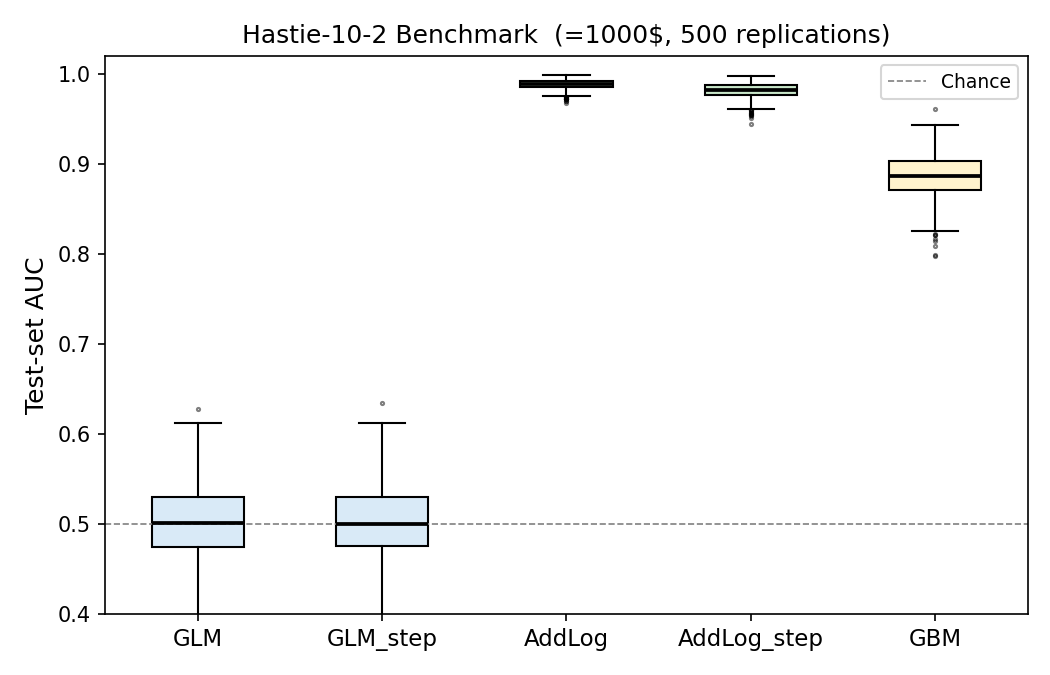}
\caption{Box-plots of test-set AUC across 500 replications on the Hastie-10-2 benchmark ($n=1000$).
The dashed horizontal line marks the chance level (AUC $= 0.5$).}
\label{fig:hastie_auc}
\end{figure}

\begin{figure}[ht]
\centering
\safeincludegraphics[width=\textwidth]{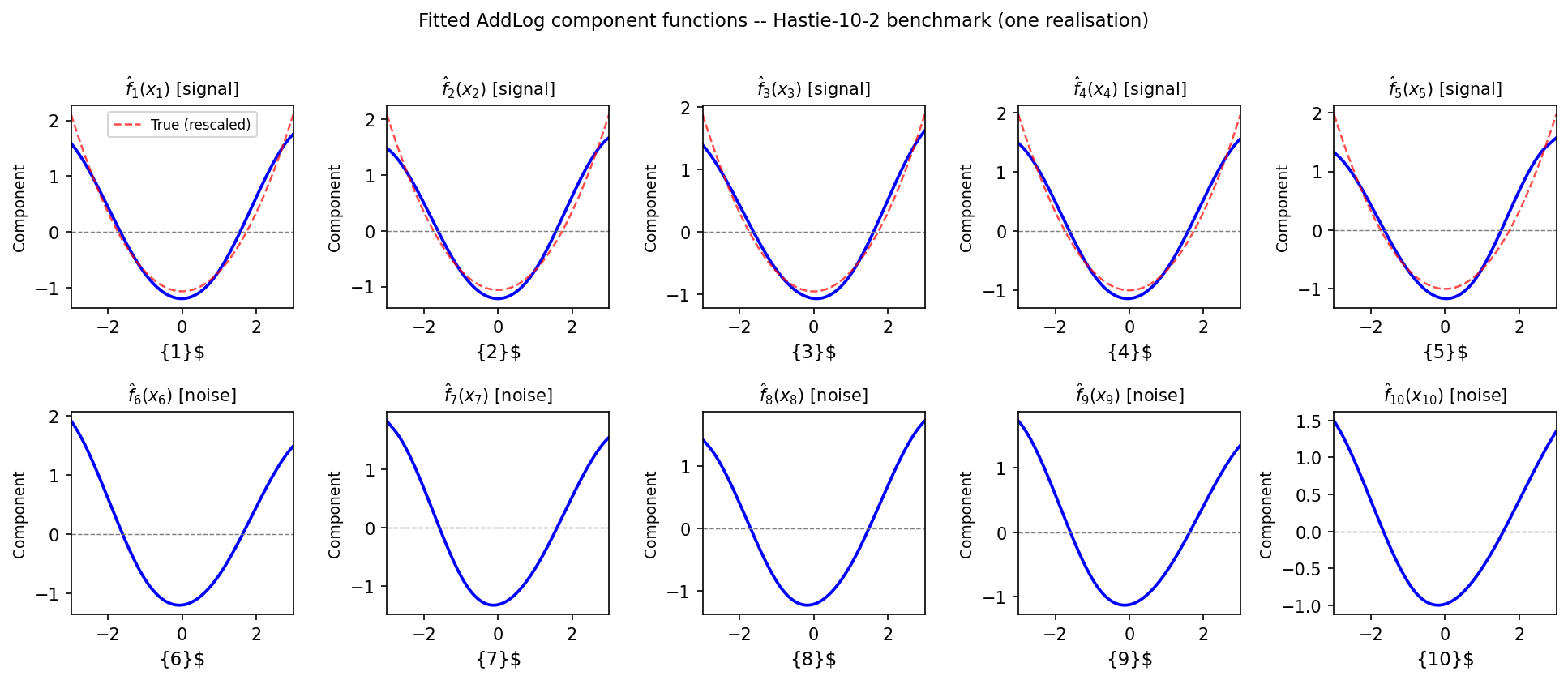}
\caption{Fitted component functions $\hat{f}_j(x_j)$ of the \textit{AddLog} model from one replication of the Hastie-10-2 benchmark.
Components $j=1,\ldots,5$ correspond to the signal predictors; components $j=6,\ldots,10$ correspond to pure noise.
The dashed red curves show the true quadratic shape $x_j^2$ (rescaled to match the fitted function's standard deviation) for the signal predictors.}
\label{fig:hastie_components}
\end{figure}

Table~\ref{tab:hastie_results} reports mean test-set AUC, its standard deviation across 500 replications, and mean computation time per replication.  Figure~\ref{fig:hastie_auc} shows the corresponding AUC box-plots, and Figure~\ref{fig:hastie_components} shows the fitted component functions of the \textit{AddLog} model from a single representative replication.

\begin{table}[htbp]
\centering
\caption{Mean test-set AUC ($\pm$ SD) and mean fitting time per replication
  across 500 replications on the Hastie-10-2 benchmark ($n=1000$, 80/20 split).
  Best AUC is shown in bold.}
\label{tab:hastie_results}
\begin{tabular}{lcccc}
\toprule
\textbf{Method} & \textbf{Mean AUC} & \textbf{SD} & \textbf{Mean time (s)} & \textbf{SD} \\
\midrule
GLM          & 0.5011 & 0.0428 & 0.0014 & 0.0001 \\
GLM\_step    & 0.5010 & 0.0431 & 0.0029 & 0.0002 \\
AddLog       & \textbf{0.9883} & 0.0055 & 0.0322 & 0.0015 \\
AddLog\_step & 0.9809 & 0.0086 & 0.1998 & 0.0379 \\
GBM          & 0.8859 & 0.0257 & 0.0845 & 0.0029 \\
\bottomrule
\end{tabular}
\end{table}

\subsection{Discussion of results}

The results illustrate the theoretical regime described in Section~\ref{sec:evidence} most clearly.
Both linear logistic methods achieve mean AUC of approximately $0.50$, confirming that no linear score can rank observations from the Hastie-10-2 distribution above chance level.
This is a direct consequence of the symmetry of the generating mechanism: the true score $\sum_{j=1}^5 x_j^2$ is orthogonal to every linear function of $x_1,\ldots,x_{10}$ under the multivariate normal distribution.
Even the $\ell_1$-penalised \textit{GLM\_step} fails to improve over \textit{GLM}, because predictor selection within the linear class cannot recover the quadratic structure.

By contrast, AddLog achieves a mean test-set AUC of 0.988 (SD = 0.006), indicating
near-perfect ranking in this benchmark.
The improvement over \textit{GLM} is approximately $0.49$ AUC units, achieved at a computational cost of $0.032$~seconds per replication compared to $0.001$~seconds for \textit{GLM}.
This represents a roughly 23-fold increase in fitting time for a near-optimal gain in ranking performance.
\textit{AddLog\_step} is slightly less accurate ($0.981$) but performs automatic noise-predictor elimination at the cost of higher computation time ($0.200$~seconds) due to the iterative SAGA solver with $\ell_1$ regularisation.

\textit{GBM} achieves a mean AUC of $0.886$, substantially above the linear models but well below the additive logistic model.
With only 50 trees of maximum depth 2, the gradient boosting machine approximates the quadratic additive boundary less efficiently than the spline basis, which directly spans the function class of the true score.

Figure~\ref{fig:hastie_components} illustrates the scientific interpretability of the \textit{AddLog} fitted model.
The five signal predictors each exhibit a clearly U-shaped (approximately quadratic) fitted component function, closely matching the true quadratic shape (dashed red curves).
The five noise predictors show flat or near-zero component functions, indicating that the model has correctly attributed no systematic effect to these variables.
This result demonstrates that an additive logistic model trained on a finite sample not only achieves near-optimal AUC, but also recovers the correct qualitative shape of each predictor's contribution to the log-odds, including the correct identification of informative versus uninformative predictors.

These findings reinforce the paper's central argument.  When the population log-likelihood ratio is nonlinear and additive, an additive logistic model can dramatically outperform a linear logistic model in terms of AUC, while providing interpretable component functions that correctly characterise the form of each predictor's effect.

% ======================================================================

\section{Summary}
This paper revisits additive logistic modelling as an interpretable approximation to the
likelihood-ratio score that underlies AUC-optimal ranking.  The main message is not that
generalized additive models are new, nor that they uniformly improve on linear logistic
regression.  Rather, the paper clarifies when nonlinear additive modelling should improve
AUC, when it should not, and what variable-level information is obtained from the fitted
component functions.

The empirical results support this conditional interpretation.  When the true or effective
ranking structure is nonlinear and approximately additive, the additive logistic score can
improve out-of-sample AUC.  When the structure is already close to linear, the additive
model may provide little or no AUC gain, but the fitted smooth functions still serve as
diagnostics of the linearity assumption.  When interactions are important, the additive
model has a clear limitation, motivating selected pairwise extensions such as GA2M.

\iffalse
This paper studies how a conventional logistic classification rule can be improved in terms of AUC while retaining variable-level interpretation.  The central message is not that a generalized additive logistic model must always dominate a linear logistic model, nor that it should replace all black-box classifiers.  Rather, the additive logistic model gives a principled intermediate option: it expands a linear log-odds score into a sum of smooth component functions, thereby allowing nonlinear threshold, saturation, and non-monotone effects to be represented without losing the ability to inspect each variable's contribution.

The results also clarify what the paper does not claim.  It does not claim that GAMs are new, nor that they always improve AUC.  It does not claim that an additive model can replace black-box methods in purely predictive settings where interpretability is unimportant.  Rather, the paper argues that additive logistic modelling remains relevant because it gives a direct, likelihood-based, variable-decomposable approximation to the AUC-optimal ranking score.  This is particularly useful in scientific and biomedical studies where the fitted model is expected to support hypothesis generation, clinical communication, and follow-up research.
\fi

At the population level, the log-likelihood-ratio score maximizes the ROC curve and AUC.  A spline-based additive logistic model can therefore improve on a conventional linear logistic model when the true ranking structure is nonlinear and approximately additive.  At the same time, the paper makes explicit what is retained and what is lost.  Marginal and quantitative variable effects remain directly interpretable through the fitted components $\hat f_j$, while unrestricted interactions are not represented unless they are added deliberately.  This clarifies the performance--interpretation trade-off: the GAM gains nonlinear ranking power relative to a GLM, while preserving much of the scientific interpretability that is typically lost in fully black-box methods.

The empirical studies support this interpretation.  In simulated examples, additive logistic models improve AUC when the data-generating mechanism contains nonlinear additive structure.  In the Parkinson's disease and cardiotocography examples, the additive model improves or matches the linear logistic model while providing component functions that can be inspected.  The additional breast cancer, SUPPORT2-calibrated critical care, and SEER-calibrated survival examples show an equally important point: when the relevant signal is already close to linear, the additive model need not improve AUC, but its fitted components still provide a diagnostic check on the adequacy of the linear assumption.  The Hastie-10-2 benchmark (Section~\ref{sec:hastie}) provides the clearest illustration of the nonlinear regime: a standard additive logistic model with cubic spline components achieves a mean test-set AUC of $0.988$ across 500 replications, compared with AUC $\approx 0.50$ for linear logistic regression.  The component function plots from this benchmark further confirm that the fitted model recovers the correct quadratic shape for signal predictors and correctly attributes near-zero effects to noise predictors.

Several numerical algorithms are available for fitting generalized additive logistic models.  The empirical results show that no single algorithm dominates in every setting.  Likelihood-based boosting is attractive for moderate sample sizes and variable-selection problems; backfitting and simultaneous smoothing-parameter estimation can be more computationally efficient for larger samples or smoother low-dimensional problems.  Practitioners should therefore treat the choice of fitting algorithm as part of the modeling strategy, especially when computation time, variable selection, and uncertainty quantification are all relevant.

Overall, the paper argues for the additive logistic model as a scientifically useful compromise between conventional logistic regression and opaque high-performance classifiers.  In applications such as biomarker research, disease diagnosis, pharmaceutical development, fetal monitoring, and clinical risk prediction, the model can improve discrimination when nonlinear effects matter, while still yielding interpretable effect functions that can motivate follow-up scientific work.

Several limitations should be emphasized.  First, the additive structure is itself an
assumption.  If the population likelihood-ratio score contains strong interactions, the
univariate component functions may give an incomplete or misleading summary.  Second,
AUC is a ranking criterion and does not by itself assess probability calibration or clinical
utility at a chosen decision threshold.  In applications where predicted probabilities are
used for decision making, calibration curves, Brier scores, and decision-analytic measures
should be reported in addition to AUC.  Third, the calibrated biomedical examples in this
paper are controlled demonstrations rather than external validation studies.  Their purpose
is to illustrate modelling behaviour, not to establish new clinical prediction rules.

\subsection{Relevance to statistical modelling}
\label{sec:relevance}

The emphasis of this paper is modelling rather than algorithmic novelty.  The additive logistic score is used to represent the log-likelihood-ratio structure that determines optimal ROC ranking.  The smooth component functions are not merely prediction devices; they are model-based summaries of how individual predictors contribute to disease risk on the log-odds scale.  This gives the approach a different role from black-box machine learning followed by post-hoc explanation.  The explanation is built into the model itself.

This modelling perspective is central to the paper's contribution.  The empirical results show when smooth marginal structure improves AUC, when it does not, and how fitted component functions can be used to understand the underlying risk structure.  The paper therefore contributes to the statistical modelling literature by connecting classical GAM methodology, ROC-based classification theory, and contemporary concerns about interpretable prediction.

The practical conclusion is that additive logistic models remain useful not because they
are a new classification method, but because they provide a transparent nonlinear scoring
model with a direct connection to likelihood-ratio ranking.  They are especially attractive
in scientific applications where the fitted classifier is expected to support both prediction
and interpretation.  Used in this way, additive logistic modelling provides a principled
middle ground between the stability of linear logistic regression and the flexibility of
opaque high-performance classifiers.

\appendix
\section{Algorithms for fitting generalized additive models}\label{Methodology}
\setcounter{equation}{0}
\renewcommand{\theequation}{A.\arabic{equation}}
\renewcommand{\theHequation}{A.\arabic{equation}}

\paragraph{Backfitting algorithm}
The backfitting algorithm employs a concept of conditional expectation; that is, if GAM is correct, then for any $k$,
%\begin{displaymath}
$E[Y-\alpha-\sum_{j\neq k}{f_j(X_j)}|X_k]=f_k(X_k).$
%\end{displaymath}
Following \eqref{gam_formula}, the backfitting algorithm is stated below;
\begin{enumerate}
	\item[(i)] Initial step: $\hat{\alpha}={N^{-1}}\sum_1^N{y_i},  \hat{f}_j\equiv0$ for $j=1,\ldots,p$.
	\item[(ii)] Cycle: $j=1,2,\ldots,p,\ldots,1,2,\ldots,p,\ldots$,
	\begin{flalign*}
	&\hat{f}_j\leftarrow S_j[\{y_i-\hat{\alpha}-\sum_{k\neq j}{\hat{f}_k(x_{ik})\}}_1^N],\\
	&\hat{f}_j\leftarrow \hat{f}_j-\frac{1}{N}\sum_{i=1}^N{\hat{f}_j(x_{ij})},
\end{flalign*}
where $S_j(\cdot)$ denotes a smoothing spline for a target against the predictor $x_j$.
	\item[(iii)] Continue Step (ii) until individual functions do not change.
\end{enumerate}

For a generalized additive logistic model as in \eqref{gam_logistic_formula}, Hastie and Tibshirani proposed a local-scoring algorithm (see Chapter 4 of their book),  
which is a maximum likelihood approach under a linear logistic regression model setup and a Newton-Raphson iterative type of method.  The local-scoring algorithm has the following steps: 
\begin{enumerate}
	\item[(i)] Initial step: $\hat{\alpha}=\log\left(\bar{y}/(1-\bar{y})\right),  \hat{f}_j\equiv0$ for $j=1,\ldots,p$.
	\item[(ii)] Cycle:
	\begin{enumerate}
		\item[(a)] Construct the current linear predictor and fitted probability,
        \[
            \hat\eta_i=\hat\alpha+\sum_{j=1}^p\hat f_j(x_{ij}),\qquad
            \hat p_i=\frac{\exp(\hat\eta_i)}{1+\exp(\hat\eta_i)}.
        \]
        The working response is
		\begin{displaymath}
			z_i=\hat\eta_i+\frac{y_i-\hat{p}_i}{\hat{p}_i(1-\hat{p}_i)}.
		\end{displaymath}
		\item[(b)] Construct weights: $w_i=\hat{p}_i(1-{\hat{p}_i})$.
		\item[(c)] Estimate the new $\hat{\alpha}$ and $\hat{f}_j$'s by fitting a weighted additive model to $z_i$ with weights $w_i$ via a backfitting algorithm.
	\end{enumerate}
	\item[(iii)] Continue Step (ii) until the individual functions do not change.
\end{enumerate}
(Note that the GAM package in R can perform both the backfitting and local scoring algorithms.)

\paragraph{Simultaneous estimation of all components with optimization in a smoothing parameter space}
Both \eqref{gam_formula} and \eqref{gam_logistic_formula} can be modeled using the $B$-spline method:
$
f_j(x)=\sum_{i=1}^K{\beta_{ij}B_{i,m}(x)},
$
where $B_{i,m}$ is the $i$th $B$-spline basis function of order $m$ for the $K+m+2$ evenly spaced knots, 
and $\beta_{ij}$ is an unknown coefficient of $B_{i,m}$ for $f_j$.  
To prevent overfitting and have a better prediction,  Marx and Eilers replaced the basic $B$-splines with a penalized $B$-splines, called $P$-splines \citep[see][]{marx1998direct}. 
Thus, when estimating the unknown $\beta_{ij}$ by maximizing a likelihood function,  
the $P$-splines will attach a different penalty on the adjacent coefficients of the $B$-splines to guarantee their smoothness. 
Therefore, to fit a GAM becomes to maximize the penalized likelihood below: 
\begin{equation}\label{Penalized-likelihood}
	l^{\star}=l(\mathbf{y};\boldsymbol{\beta})-\frac{1}{2}\sum_{j=1}^p{\lambda_j\boldsymbol{\beta}_j^T\mathbf{P}\boldsymbol{\beta}_j},
\end{equation} 
where the $\lambda_j$'s are smoothing parameters,  $\mathbf{P}$ is the penalty matrix, 
and $l(\mathbf{y};\boldsymbol{\beta})$ is the log-likelihood of GAM. 
For a  generalized additive logistic model as  in \eqref{gam_logistic_formula}, the log-likelihood with a penalized first order differences becomes
\begin{displaymath}
	l^{\star}=l(\mathbf{y};\boldsymbol{\beta})-\frac{1}{2}\sum_{j=1}^p{\left\{\lambda_j\sum_{i=1}^{K-1}{(\beta_{i+1,j}-\beta_{i,j})^2}\right\}},
\end{displaymath}
where $l(\mathbf{y};\boldsymbol{\beta})=\sum_{i=1}^n{\{y_i\log{p_i}+(1-y_i)\log{(1-p_i)}\}}$.
\citet{wood2004stable} provided a stable and efficient approach to estimate all functions, simultaneously. This feature is available in the \emph{mgcv} package in R. 

\paragraph{Likelihood-based boosting method}
%Boosting is a powerful technology and has been successfully derived for improving classification problems. 
Its basic idea of the boosting algorithm is to combine some ``weaker classifiers'' to produce a more powerful classifier 
that can reduce misclassification error using an iterative adaptive weighting scheme on observations. 
\citet{tutz2006generalized} use it as a way of fitting an additive expansion in a set of elementary basis functions  by minimizing a loss function,  such as a likelihood-based loss function.  
They proposed the  GAMBoost procedure to fit GAM with this boosting technology via maximizing the log-likelihood \eqref{Penalized-likelihood}.  
In the statements below, we also use similar notations as that in the backfitting. 
The GAMBoost algorithm for fitting a generalized additive logistic model \eqref{gam_logistic_formula} can be summarized as follows:
\begin{enumerate}
	\item[(i)] Define the additive predictor and the fitted mean by
    \[
        \eta_i=\alpha+\sum_{j=1}^p f_j(x_{ij}),\qquad
        \mu_i=h(\eta_i)=\frac{\exp(\eta_i)}{1+\exp(\eta_i)}.
    \]
	\item[(ii)] Initialize
    \[
        \hat\alpha=\log\left(\frac{\bar y}{1-\bar y}\right),\qquad
        \hat f_j^{(0)}\equiv 0,
    \]
    so that $\hat\eta_i^{(0)}=\hat\alpha$ and $\hat\mu_i^{(0)}=h(\hat\eta_i^{(0)})$.
	\item[(iii)] For boosting step $m=0,1,2,\ldots$:
	\begin{enumerate}
	\item[(a)] For each component $j$, form the basis matrix $\mathbf B_j$ with $i$th row
    \[
        \mathbf B_{ij}=\{B_{1,m}(x_{ij}),\ldots,B_{K,m}(x_{ij})\},
    \]
    and compute the working residual and weights
    \[
        r_i^{(m)}=\frac{y_i-\hat\mu_i^{(m)}}{\hat\mu_i^{(m)}(1-\hat\mu_i^{(m)})},\qquad
        w_i^{(m)}=\hat\mu_i^{(m)}(1-\hat\mu_i^{(m)}).
    \]
    With $\mathbf W^{(m)}=\operatorname{diag}(w_1^{(m)},\ldots,w_n^{(m)})$, estimate the candidate coefficient vector
    \[
        \hat{\mathbf b}_j^{(m)}=
        (\mathbf B_j^T\mathbf W^{(m)}\mathbf B_j+\lambda_j\mathbf P_j)^{-1}
        \mathbf B_j^T\mathbf W^{(m)}\mathbf r^{(m)}.
    \]
    The candidate update for component $j$ is $\mathbf u_j^{(m)}=\mathbf B_j\hat{\mathbf b}_j^{(m)}$.
	\item[(b)] Select the component that yields the largest decrease in binomial deviance:
    \[
        s_m=\arg\min_{1\le j\le p}\operatorname{Dev}\{\hat{\boldsymbol\eta}^{(m)}+\mathbf u_j^{(m)}\}.
    \]
	\item[(c)] With a step length $0<\nu\le 1$, update
    \[
        \hat{\boldsymbol\eta}^{(m+1)}=\hat{\boldsymbol\eta}^{(m)}+\nu\mathbf u_{s_m}^{(m)},
        \qquad
        \hat f_{s_m}^{(m+1)}=\hat f_{s_m}^{(m)}+\nu u_{s_m}^{(m)}.
    \]
    All other components are unchanged.
	\end{enumerate}
\end{enumerate}
The R package {GAMBoost} is developed based on this type of componentwise likelihood-based boosting algorithm.

\paragraph{Empirical AUC-contribution summaries}

For an additive logistic score
\[
\hat S(x)=\hat\alpha+\sum_{j=1}^p \hat f_j(x_j),
\]
we compute two optional diagnostic summaries of variable contribution to AUC.

The deletion contribution is
\[
\Delta^{\mathrm{del}}_j
=
\widehat{\mathrm{AUC}}\{\hat S(X)\}
-
\widehat{\mathrm{AUC}}\{\hat S_{-j}(X)\},
\qquad
\hat S_{-j}(x)=\hat\alpha+\sum_{k\neq j}\hat f_k(x_k).
\]

The permutation contribution is
\[
\Delta^{\mathrm{perm}}_j
=
\widehat{\mathrm{AUC}}\{\hat S(X)\}
-
\widehat{\mathrm{AUC}}\{\hat S(X_{-j},X^{\pi}_j)\},
\]
where \(X^{\pi}_j\) is obtained by randomly permuting predictor \(j\) in the test set.
The permutation is repeated \(B\) times and the contributions are averaged.  Bootstrap
intervals can be obtained by resampling the test set and recomputing the AUC differences.
These summaries are diagnostic and should not be interpreted as a unique population
decomposition of AUC.


\begin{thebibliography}{99}
\expandafter\ifx\csname natexlab\endcsname\relax\def\natexlab#1{#1}\fi
\expandafter\ifx\csname url\endcsname\relax\def\url#1{\texttt{#1}}\fi
\expandafter\ifx\csname urlprefix\endcsname\relax\def\urlprefix{URL }\fi

\bibitem[Benjamini and Hochberg(1995)]{benjamini1995controlling}
Benjamini, Y. and Hochberg, Y. (1995).
\newblock Controlling the false discovery rate: a practical and powerful
  approach to multiple testing.
\newblock \textit{Journal of the Royal Statistical Society: Series B},
  57(1):289--300.

\bibitem[Devroye et~al.(1996)]{devroye1996probabilistic}
Devroye, L., Gy\"{o}rfi, L., and Lugosi, G. (1996).
\newblock \textit{A Probabilistic Theory of Pattern Recognition}.
\newblock Springer, New York.

\bibitem[Friedman et~al.(2000)]{friedman2000additive}
Friedman, J., Hastie, T., and Tibshirani, R. (2000).
\newblock Additive logistic regression: a statistical view of boosting.
\newblock \textit{Annals of Statistics}, 28(2):337--407.

\bibitem[Hastie and Tibshirani(1990)]{hastie1990generalized}
Hastie, T. and Tibshirani, R. (1990).
\newblock \textit{Generalized Additive Models}.
\newblock Chapman \& Hall, London.

\bibitem[Pepe(2004)]{pepe2004statistical}
Pepe, M.~S. (2004).
\newblock \textit{The Statistical Evaluation of Medical Tests for
  Classification and Prediction}.
\newblock Oxford University Press.

\bibitem[Wood(2006)]{wood2006generalized}
Wood, S.~N. (2006).
\newblock \textit{Generalized Additive Models: An Introduction with R}.
\newblock Chapman \& Hall/CRC, Boca Raton.

\bibitem[{Bache and Lichman(2013)}]{Bache+Lichman:2013}
Bache, K., Lichman, M., 2013. {UCI} machine learning repository.
\newline\urlprefix\url{http://archive.ics.uci.edu/ml}

\bibitem[{Binder and Tutz(2008)}]{binder-tutz08}
Binder, H., Tutz, G., 2008. A comparison of methods for the fitting of
  generalized additive models. Stat. Comput 18, 87 -- 99.

\bibitem[{Cristianini and Shawe-Taylor(2000)}]{SVM-intro}
Cristianini, N., Shawe-Taylor, J., 2000. An Introduction to Support Vector
  Machines and Other Kernel-based Learning Methods. Cambridge University Press,
  The Edinburgh Building, Cambridge CB2 8RU, UK.

\bibitem[{Eguchi and Copas(2002)}]{eguchi2002class}
Eguchi, S., Copas, J., 2002. A class of logistic-type discriminant functions.
  Biometrika 89~(1), 1--22.

\bibitem[{Friedman et~al.(2000)Friedman, Hastie, and Tibshirani}]{Friedman2k}
Friedman, J., Hastie, T., Tibshirani, R., 2000. Additive logistic regression: a
  statistical view of boosting. Ann. Statist., 337 -- 407.

\bibitem[{Hastie and Tibshirani(1986)}]{hastie}
Hastie, T., Tibshirani, R., 1986. Generalized additive models. Statistical
  Science 1~(3), 297 -- 318.

\bibitem[{Hastie and Tibshirani(1990)}]{GAM}
Hastie, T., Tibshirani, R., 1990. Generalized additive models. Vol.~43. CRC
  Press.

\bibitem[{Hastie et~al.(2001)Hastie, Tibshirani, and
  Friedman}]{hastie-tibshirani-friedman2001}
Hastie, T., Tibshirani, R., Friedman, J., 2001. The Elements of Statistical
  Learning. Springer, Berlin.

\bibitem[{Little et~al.(2007)Little, McSharry, Roberts, Costello, Moroz,
  et~al.}]{little2007exploiting}
Little, M.~A., McSharry, P.~E., Roberts, S.~J., Costello, D.~A., Moroz, I.~M.,
  et~al., 2007. Exploiting nonlinear recurrence and fractal scaling properties
  for voice disorder detection. BioMedical Engineering OnLine 6~(1), 23.

\bibitem[{Marra and Wood(2012)}]{marra-wood12}
Marra, G., Wood, S., 2012. Coverage properties of confidence intervals for
  generalized additive model components. Scandinavian Journal of Statistics
  39~(1), 53 --74.

\bibitem[{Marx and Eilers(1998)}]{marx1998direct}
Marx, B.~D., Eilers, P.~H., 1998. Direct generalized additive modeling with
  penalized likelihood. Computational Statistics \& Data Analysis 28~(2),
  193--209.

\bibitem[{Minka(2007)}]{Minka07}
Minka, T., March 26 2007. A comparison of numerical optimizers for logistic
  regression.
\newline\urlprefix\url{http://research.microsoft.com/en-us/um/people/minka/papers/logreg/}

\bibitem[{Stone(1985)}]{stone1985additive}
Stone, C.~J., 1985. Additive regression and other nonparametric models. The
  annals of Statistics, 689--705.

\bibitem[{Tian et~al.(2007)Tian, Cai, Goetghebeur, and Wei}]{Tian07}
Tian, L., Cai, T., Goetghebeur, E., Wei, L.~J., 2007. Model evaluation based on
  the sampling distribution of estimated absolute prediction error. Biometrika
  94~(2), 297--311.

\bibitem[{Tutz and Binder(2006)}]{tutz2006generalized}
Tutz, G., Binder, H., 2006. Generalized additive modeling with implicit
  variable selection by likelihood-based boosting. Biometrics 62~(4), 961--971.

\bibitem[{Wood(2004)}]{wood2004stable}
Wood, S., 2004. Stable and efficient multiple smoothing parameter estimation
  for generalized additive models. Journal of the American Statistical
  Association 99~(467).

\bibitem[{Wood(2006)}]{Wood06}
Wood, S., 2006. Generalized Additive Models: An Introduction with R. Chapman \&
  Hall.

\bibitem[{Wood(2008)}]{Wood08}
Wood, S., 2008. Fast stable direct fitting and smoothness selectionfor
  generalized additive models. Journal of Royal Statist. Soc. B. 70~(3), 495 --
  518.

\bibitem[{Knaus et al.(1995)}]{knaus1995support}
Knaus, W.~A., Harrell, F.~E., Lynn, J., Goldman, L., Phillips, R.~S.,
  Connors, A.~F., Dawson, N.~V., Fulkerson, W.~J., Califf, R.~M., Desbiens,
  N., et~al., 1995. The {SUPPORT} prognostic model: Objective estimates of
  survival for seriously ill hospitalized adults. Annals of Internal Medicine
  122~(3), 191--203.

\bibitem[{National Cancer Institute(2022)}]{seer2022}
National Cancer Institute, 2022. {SEER} Cancer Statistics Review,
  1975--2020. Surveillance, Epidemiology, and End Results Program.
\newline\urlprefix\url{https://seer.cancer.gov/csr/1975_2020/}

\bibitem[{Lou et~al.(2012)Lou, Caruana, and Gehrke}]{lou2012intelligible}
Lou, Y., Caruana, R., Gehrke, J., 2012. Intelligible models for classification and regression.
  In: \textit{Proceedings of the 18th ACM SIGKDD International Conference on Knowledge Discovery and Data Mining}, pp.~150--158.

\bibitem[{Lou et~al.(2013)Lou, Caruana, Gehrke, and Hooker}]{lou2013accurate}
Lou, Y., Caruana, R., Gehrke, J., Hooker, G., 2013. Accurate intelligible models with pairwise interactions.
  In: \textit{Proceedings of the 19th ACM SIGKDD International Conference on Knowledge Discovery and Data Mining}, pp.~623--631.

\bibitem[{Caruana et~al.(2015)Caruana, Lou, Gehrke, Koch, Sturm, and Elhadad}]{caruana2015intelligible}
Caruana, R., Lou, Y., Gehrke, J., Koch, P., Sturm, M., Elhadad, N., 2015. Intelligible models for healthcare: Predicting pneumonia risk and hospital 30-day readmission.
  In: \textit{Proceedings of the 21st ACM SIGKDD International Conference on Knowledge Discovery and Data Mining}, pp.~1721--1730.

\bibitem[{Agarwal et~al.(2021)Agarwal, Melnick, Frosst, Zhang, Lengerich, Caruana, and Hinton}]{agarwal2021nam}
Agarwal, R., Melnick, L., Frosst, N., Zhang, X., Lengerich, B., Caruana, R., Hinton, G.~E., 2021. Neural additive models: Interpretable machine learning with neural nets.
  In: \textit{Advances in Neural Information Processing Systems}, Vol.~34, pp.~4699--4711.

\bibitem[{Rudin(2019)}]{rudin2019stop}
Rudin, C., 2019. Stop explaining black box machine learning models for high stakes decisions and use interpretable models instead.
  Nature Machine Intelligence 1, 206--215.

\bibitem[{Lundberg and Lee(2017)}]{lundberg2017shap}
Lundberg, S.~M., Lee, S.-I., 2017. A unified approach to interpreting model predictions.
  In: \textit{Advances in Neural Information Processing Systems}.

\bibitem[{Goldstein et~al.(2015)Goldstein, Kapelner, Bleich, and Pitkin}]{goldstein2015ice}
Goldstein, A., Kapelner, A., Bleich, J., Pitkin, E., 2015. Peeking inside the black box: Visualizing statistical learning with plots of individual conditional expectation.
  Journal of Computational and Graphical Statistics 24~(1), 44--65.

\bibitem[{Vickers and Elkin(2006)}]{vickers2006decision}
Vickers, A.~J., Elkin, E.~B., 2006. Decision curve analysis: A novel method for evaluating prediction models.
  Medical Decision Making 26~(6), 565--574.

\end{thebibliography}
\end{document}